\documentclass[aps,pra,reprint,amsmath,amssymb]{revtex4-2}
\usepackage[utf8]{inputenc}
\usepackage[T1]{fontenc}
\usepackage{hyperref}
\usepackage{graphicx}
\usepackage[normalem]{ulem}
\usepackage{bbm}
\usepackage{bm}
\usepackage[caption=false]{subfig}
\usepackage{color}
\usepackage{dsfont}
\usepackage{bigints}

\DeclareMathOperator{\tr}{tr}

\newcommand{\phid}{\phi^\dag}
\newcommand{\psid}{\psi^\dag}

\newcommand{\VPHI}{\bm{\varphi}}
\newcommand{\PHI}{\bm{\phi}}

\newcommand{\MR}{\mathbf{R}}
\newcommand{\MG}{\mathbf{G}}
\newcommand{\MGamma}{\bm{\Gamma}}
\newcommand{\Q}{\mathbf{q}}
\newcommand{\X}{\mathbf{x}}
\newcommand{\vP}{\mathbf{p}}
\newcommand{\ddX}{\mathrm{d}^d\X}

\newcommand{\ddQ}{\mathrm{d}^d\Q}
\newcommand{\dtau}{\mathrm{d} \tau}

\begin{document}

\title{Functional renormalization for repulsive Bose-Bose mixtures at zero temperature}
\author{Felipe Isaule}
\author{Ivan Morera}
\author{Artur Polls}
\author{Bruno Juli\'{a}-D\'{i}az}
\affiliation{Departament de F\'isica Qu\`antica i Astrof\'isica, 
Facultat de F\'{\i}sica, and Institut de Ci\`encies del Cosmos (ICCUB), Universitat de Barcelona, 
Mart\'i i Franqu\`es 1, E–08028 Barcelona, Spain}
\date{\today}

\begin{abstract}

We study weakly-repulsive Bose-Bose mixtures in two and three dimensions at zero temperature 
using the functional renormalization group (FRG). We examine the RG flows and the role of density
and spin fluctuations. We study the condition for phase separation and find that this occurs
at the mean-field point within the range of parameters explored. Finally, we examine
the energy per particle and condensation depletion. We obtain that our FRG calculations
compare favorably with known results from perturbative approaches for macroscopic properties.

\end{abstract}

\maketitle

\section{Introduction}

Weakly-interacting Bose gases have been subject of study for many decades~\cite{pethick_bose-einstein_2008,pitaevskii_bose-einstein_2016}.
However, the interest on such gases greatly increased since the experimental 
realization of Bose-Einstein condensation with cold alkali atoms~\cite{anderson_observation_1995,davis_bose-einstein_1995,bradley_evidence_1995}.

Experimentalists have been able to produce bosonic cold-atom gases in a variety 
of configurations and explore a range of interaction parameters 
and temperatures~\cite{bloch_many-body_2008}. Theoretically, even though 
mean-field (MF) theory is able to give a qualitative description of Bose 
gases at low temperatures~\cite{bogoliubov_n_theory_1947}, for an accurate 
description, the effect of fluctuations needs to be included. For one-component 
Bose gases, the leading quantum correction in three dimensions was first 
calculated by Lee, Huang, and Yang (LHY)~\cite{lee_many-body_1957,lee_eigenvalues_1957}. 
Since then, further improvements have been provided with a variety 
of approaches~\cite{andersen_theory_2004}. In one and two dimensions, the effect 
of fluctuations is enhanced~\cite{al_khawaja_low_2002}. Thus, corrections beyond 
the zero-point level in perturbation theory are more important, often 
requiring more careful treatments~\cite{posazhennikova__2006,hadzibabic_two-dimensional_2011,cazalilla_one_2011}. 
In general, nowadays, one-component Bose gases are considered well described, 
and therefore the interest has shifted towards more sophisticated related systems. 

Bose-Bose mixtures, gases with two species of bosons~\cite{pitaevskii_bose-einstein_2016},
have attracted attention in recent years. The interplay between the two components 
of the gas leads to rich physics such as the superfluid 
drag~\cite{andreev_three-velocity_1975,fil_nondissipative_2005,nespolo_andreevbashkin_2017}, 
Josephson effect~\cite{tommasini_bogoliubov_2003,abad_study_2013} and, depending if 
the inter-species interaction is repulsive or attractive, phase 
separation~\cite{ao_binary_1998,timmermans_phase_1998,trippenbach_structure_2000} 
and self-bound droplets~\cite{petrov_quantum_2015,petrov_ultradilute_2016}.
Mixtures of bosons in cold atom gases were rapidly achieved experimentally by 
using bosonic atoms with two hyperfine states~\cite{ho_binary_1996,myatt_production_1997,hall_dynamics_1998,hall_measurements_1998,modugno_two_2002}, and there is currently a great effort on producing droplet 
phases in different configurations~\cite{chomaz_quantum-fluctuation-driven_2016,ferrier-barbut_observation_2016,cappellaro_equation_2017,cabrera_quantum_2018}.
Theoretically, Bose-Bose mixtures have been studied with different techniques, 
most noticeable perturbative approaches~\cite{larsen_binary_1963,oles_n_2008,armaitis_hydrodynamic_2015,armaitis_hydrodynamic_2015,chiquillo_equation_2018,konietin_2d_2018,ota_beyond_2020},
Beliaev theory~\cite{utesov_effective_2018} and Monte Carlo (MC) simulations~\cite{petrov_ultradilute_2016,parisi_liquid_2019,cikojevic_universality_2019}.

One alternative theoretical approach used to study quantum gases is the functional
renormalization group (FRG) based on the effective average action~\cite{wetterich_exact_1993,berges_non-perturbative_2002}.
This is a non-perturbative technique where the full effective action of the 
system is calculated by integrating out fluctuations, both quantum and thermal, 
by means of an RG equation. The FRG has been used with great success to study 
weakly-interacting one-component Bose gases in different dimensions~\cite{dupuis_non-perturbative_2007,wetterich_functional_2008,floerchinger_functional_2008,floerchinger_superfluid_2009,floerchinger_nonperturbative_2009,dupuis_infrared_2009,sinner_functional_2010,lammers_dimensional_2016,isaule_thermodynamics_2020}.
One advantage of this approach is that the intricate long-distance physics is 
gradually incorporated during the RG flow~\cite{dupuis_infrared_2011}. Therefore 
the FRG does not suffer from the infrared (IR) divergences that plague 
perturbation theory~\cite{gavoret_structure_1964,nepomnyashchii_contribution_1975}.
Furthermore, as a non-perturbative technique, the FRG has proved to deal with
strongly-interacting systems with ease. For Bose gases, this has been relevant 
to study strongly-correlated superfluids in optical lattices~\cite{rancon_nonperturbative_2011}.

Building upon the works on one-component gases, we study balanced and weakly-repulsive Bose-Bose mixtures at zero temperature within the FRG . We 
focus on both the RG flows and thermodynamics in two and three dimensions.
We also examine the condition for phase separation.
These configurations have been recently studied in detail using perturbative approaches 
in Refs.~\cite{armaitis_hydrodynamic_2015,chiquillo_equation_2018,konietin_2d_2018,ota_beyond_2020}.
RG approaches have also been used to study Bose-Bose mixtures. However, such studies
have focused exclusively on the phase separation around the zero-density critical point~\cite{kolezhuk_stability_2010,lee_stability_2011}
and the superfluid phase transition~\cite{ceccarelli_bose-einstein_2015,karle_coupled_2019}.

This work is organized in the following way. First, in Sec.~\ref{sec:S},
we present the microscopic model within the path-integral formalism.
In Sec.~\ref{sec:FRG}, we present the FRG formalism and the ansatz for
the effective action, including a short discussion on the physical
inputs and the momentum regimes.
The results are presented in Sec.~\ref{sec:results}, where
we examine some features of the RG flows, the phase separation point,
and finally, some thermodynamic properties.

\section{Microscopic model}
\label{sec:S}

We consider a non-relativistic gas of two species of bosons, $A$ and $B$, 
interacting through weak repulsive inter- and intra-species interactions.
We work at length scales where the microscopic details of the interactions 
are not important, which are therefore represented by effective contact 
potentials with strength $g_{ab}$. We study the 
balanced mixture, that is, both species of bosons have equal masses $m=m_A=m_B$, 
equal chemical potentials $\mu=\mu_A=\mu_B$ and equal intra-species 
interactions $g=g_{AA}=g_{BB}$.

In a path-integral formulation~\cite{stoof_ultracold_2009} such gas is described 
by the Euclidean microscopic action
\begin{multline}
\mathcal{S}[\VPHI]=\int_x\sum_{a=A,B}
\Bigg[\psid_a\left(\partial_\tau-\frac{\nabla^2}{2m}-\mu\right)\psi_a\\
+\sum_{b=A,B}\frac{g_{ab}}{2}|\psi_a|^2|\psi_b|^2\Bigg],
\label{sec:S;eq:S}
\end{multline}
where $\int_x=\int_0^\beta\dtau\int\ddX$, with $\tau=it$ the imaginary time and 
$\beta=T^{-1}$ the inverse temperature. In this work we study the zero-temperature 
gas, hence $\beta\to\infty$. The action is a functional of the fields
$\VPHI=(\psi_A,\psid_A,\psi_B,\psid_B)$, which represent the two bosonic species.
Note that we set $\hbar=k_B=1$. 

Throughout this article, we work in Fourier space $q=(\omega,\Q)$. The kinetic terms in the action take the form
\begin{equation}
    \mathcal{S}_\text{kin}[\VPHI]=\int_q\sum_{a=A,B}
    \left[\psid_a\left(i\omega+\frac{\Q^2}{2m}\right)\psi_a\right]\,,
\end{equation}
where the integral over $q$ at zero temperature is
\begin{equation}
 \int_q=\int_{-\infty}^{\infty}\frac{\mathrm{d}\omega}{2\pi}\int\frac{\ddQ}{(2\pi)^d}\,.
\label{sec:S;eq:intq}
\end{equation}
The microscopic action defines the grand-canonical partition function
\begin{equation}
\mathcal{Z}[\VPHI]=\int D \VPHI e^{-\mathcal{S}[\VPHI]}\,,
\end{equation}
which is a \emph{path integral} over all configurations of the fields. 
The associated thermodynamic potential is obtained through
\begin{equation}
 \Omega=-\beta^{-1}\log \mathcal{Z}\,,
\end{equation}
from which we can extract all the thermodynamic properties of the system.
In the balanced mixture, its differential at zero temperature is
\begin{equation}
 \mathrm{d}\Omega=-P\mathrm{d}\mathcal{V}_d-\langle N \rangle \mathrm{d}\mu\,,
\end{equation}
where $\mathcal{V}_d$ is the $d$-dimensional volume, $P$ the pressure and
$N$ the total number of particles. The energy density of the ground state is
\begin{equation}
 \frac{E}{\mathcal{V}_d}=-P \mathcal+ n \mu\,,
\label{sec:S;eq:E}
\end{equation}
where $n=n_A+n_B$ is the total atom density. The energy per particle 
is obtained from $E/N=(E/\mathcal{V}_d)/n$.

\section{Functional renormalization group}
\label{sec:FRG}

The strategy of the FRG is to generate the \emph{effective action} $\Gamma$ 
of a system by smoothly taking fluctuations into account with an RG 
group equation~\cite{wetterich_exact_1993,berges_non-perturbative_2002}. To achieve 
this, one considers a regulator function $\MR(q;k)$, which suppresses 
all fluctuations for momenta $q\lesssim k$. This regulator is added to the theory as a 
mass term
\begin{equation}
\Delta \mathcal{S}_k[\VPHI]=\int_q\VPHI^\dag(q)\MR_k(q)\VPHI(q)\,,
\end{equation}
so the grand-canonical partition function becomes $k$-dependent
\begin{equation}
\mathcal{Z}_k[\mathbf{J}]=\int D \VPHI e^{-\mathcal{S}[\VPHI]
-\Delta \mathcal{S}_k[\VPHI]+\int_x \mathbf{J}\cdot\VPHI}\,,
\end{equation}
where we also added source fields $\mathbf{J}$. The $k$-dependent effective 
action is defined by means of a Legendre transformation
\begin{equation}
\Gamma_k[\PHI]=-\log\mathcal{Z}_k[\mathbf{J}]
+\int_x \mathbf{J}\cdot\PHI-\Delta\mathcal{S}_k[\PHI],
\end{equation}
where $\PHI(x)=\langle\VPHI(x)\rangle$ are classical fields. At a UV 
scale $k=\Lambda$, all fluctuations are suppressed and the 
effective action is simply the microscopic action; $\Gamma_\Lambda[\PHI]=\mathcal{S}[\PHI]$.
In contrast, for $k\to 0$, all fluctuations have been taken into account and 
$\Gamma_0[\PHI]$ is the physical effective action. $\Gamma_0$ is the generator 
of the one-particle irreducible Green’s functions, from which we can extract 
the physical properties of interest. At equilibrium
\begin{equation}
    \frac{\delta\Gamma}{\delta\PHI}\bigg|_{\PHI_0}=0,
\end{equation}
the effective action is related to the thermodynamic potential $\Omega$ through
\begin{equation}
 \Omega=\Gamma[\PHI_0]/\beta\,,
\end{equation}
which enables us to extract the thermodynamic properties of the gas.

The flow of $\Gamma_k$ as a function of $k$ is dictated by the Wetterich
equation~\cite{wetterich_exact_1993}
\begin{equation}
\partial_k \Gamma_k=\frac{1}{2}\tr\left[\partial_k \MR_k(\MGamma_k^{(2)}+\MR_k)^{-1}\right]\,,
\label{sec:FRG;eq:Wetterich}
\end{equation}
where $\MGamma^{(2)}_k$ is the matrix with the second-functional 
derivatives of $\Gamma_k$. The trace denotes both a trace and an 
integral over internal momentum $q$.

In most applications, the Wetterich equation cannot be solved directly, and 
thus approximations need to be employed. In this work, we propose an ansatz 
for the effective action $\Gamma_k$ based on a derivative expansion (DE) 
truncated to a small number of $k$-dependent terms~\cite{berges_non-perturbative_2002}. 
Within this approximation, the Wetterich equation becomes a 
set of coupled differential equations that can be solved numerically 
using standard methods. The DE is generally appropriate to study the 
long-distance physics and thermodynamic properties as in this work.

In the following we present the ansatz for the effective action and 
details of the flow equations. For detailed reviews on the FRG framework 
we refer to Refs.~\cite{berges_non-perturbative_2002,delamotte_nonperturbative_2004,kopietz_introduction_2010,boettcher_ultracold_2012,dupuis_nonperturbative_2020}.

\subsection{Effective action}
\label{sec:FRG;sub:ansatz}

We use an ansatz and strategy analogous to those used for one-component 
Bose gases~\cite{dupuis_non-perturbative_2007,wetterich_functional_2008,floerchinger_functional_2008}. Our ansatz  reads
\begin{multline}
\Gamma_k[\PHI]=\int_x\bigg[\sum_{a=A,B}\psid_a\left(S\partial_\tau-\frac{Z}{2m}\nabla^2
-V\partial_\tau^2\right)\psi_a\\
+U(\rho_A,\rho_B;\mu)\bigg]\,,
\label{sec:FRG;sub:ansatz;eq:Gamma}
\end{multline}
where $\rho_a=\phid_a\phi_a$ and $S$, $Z$ and $V$ are $k$-dependent 
renormalization factors, which we consider as field-independent. 
The effective potential $U$ is expanded around 
the $k$-dependent order parameters $\rho_{a,0}=\langle\phid_a\phi_a\rangle$, 
which are equal for the balanced mixture $\rho_0=\rho_{A,0}=\rho_{B,0}$ .
We are interested in the phase where the $U(1)$-symmetry of each species 
is broken, thus $\rho_0>0$ for \emph{all} $k$~\cite{floerchinger_functional_2008}. 
In this phase, the effective potential truncated up to 
quartic order in the fields reads
\begin{align}
U(\rho_A,\rho_B;\mu)=&-\sum_{a=A,B}
(n_0+n_1(\rho_a-\rho_0))(\tilde{\mu}-\mu) \nonumber\\
&+\sum_{a,b=A,B}\frac{\lambda_{ab}}{2}(\rho_a-\rho_0)(\rho_b-\rho_0),
\end{align}
where $\lambda_{AA}=\lambda_{BB}=\lambda$.
We have added shifts $\tilde{\mu}$ from the from the physical chemical potential 
$\mu$, so we can follow the flow of the densities $n_0=n/2$ of each species 
of bosons (see Ref.~\cite{floerchinger_functional_2008} for details). The 
condensate densities are given by $\rho_0$, whereas the superfluid densities 
are given by the phase stiffness 
$\rho_s=Z\rho_0$~\cite{popov_functional_1987,dupuis_infrared_2009}.
Thus, the physical condensate and superfluid fractions are obtained from 
the values at $k=0$ of $\Omega_s=\rho_s/n_0$ and $\Omega_c=\rho_0/n_0$, respectively.

When the $U(1)$-symmetry is broken, it is useful to decompose the complex 
boson fields $\psi_a$ into orthogonal real fields
\begin{equation}
 \psi_a(x)=\frac{1}{\sqrt{2}}\left(\psi_{a,1}(x)+i\psi_{a,2}(x)\right)\,.
 \label{sec:FRG;sub:ansatz;eq:psi12}
\end{equation}
In this work we fix both order parameters
$\rho_{A,0}$ and $\rho_{B,0}$ at the same direction. Thus, we evaluate the 
fields at $\langle\psi_{a,1}\rangle=(2\rho_0)^{1/2}$ and $\langle\psi_{a,2}\rangle=0$.
The inverse propagator evaluated at this point reads
\begin{equation}
  \MG^{-1}_k(q)=\begin{pmatrix}
              \MG^{-1}_{k,\psi}(q) & \bm{\Sigma}_{k,AB}\\
              \bm{\Sigma}_{k,BA} &  \MG^{-1}_{k,\psi}(q)
             \end{pmatrix}\,.
             \label{sec:FRG;sub:ansatz;eq:Ginv}
\end{equation}
where
\begin{equation}
  \MG^{-1}_{k,\psi}(q)=\begin{pmatrix}
              E_{1,k}(\Q)+V\omega & S \omega\\
              -S \omega & E_{2,k}(\Q)+V\omega
             \end{pmatrix}\,,
\end{equation}
\begin{equation}
  \bm{\Sigma}_{k,AB}=\bm{\Sigma}_{k,BA}=\begin{pmatrix}
              2\lambda_{AB}\rho_0 & 0\\
              0 &  0
             \end{pmatrix},
\end{equation}
and
\begin{align}
 E_{1,k}(\Q)=&Z\frac{\Q^2}{2m}+2\lambda\rho_0+n_1(\tilde{\mu}-\mu)+R_k(\Q), \label{sec:FRG;sub:ansatz;eq:E1}\\
 E_{2,k}(\Q)=&Z\frac{\Q^2}{2m}+n_1(\tilde{\mu}-\mu)+R_k(\Q). \label{sec:FRG;sub:ansatz;eq:E2}
\end{align}
As with one-component gases, $\psi_{a,1}$ represent fluctuations
of the longitudinal modes, whereas $\psi_{a,2}$ represent fluctuations of 
the massless Goldstone modes~\cite{wetterich_functional_2008}.

In this work we use the optimized Litim regulator~\cite{litim_optimisation_2000}, 
\begin{equation}
 R_k(\Q)=\frac{Z}{2m}(k^2-\Q^2)\Theta(\Q^2-k^2)\,,
\label{sec:FRG;sub:flow_eqs;eq:Litim}
\end{equation}
where $\Theta$ is the Heaviside step function. This regulator allows 
us to perform the momentum integrals analytically before solving the 
differential equations. Although this regulator is frequency-independent, it 
has proven to give reasonable results for 
quantum gases~\cite{floerchinger_nonperturbative_2009,boettcher_ultracold_2012}.

The flow equations for the $k$-dependent are obtained from the
appropriate projections of the Wetterich equation~(\ref{sec:FRG;eq:Wetterich}) into the ansatz~(\ref{sec:FRG;sub:ansatz;eq:Gamma}), which
are analogous to those for one-component gases~\cite{floerchinger_functional_2008,isaule_thermodynamics_2020}.
These can be found in App.~\ref{app:flow_eqs}.

We stress that in our ansatz, we do not include the effect of 
the relative phase between the two condensates, and therefore 
we do not describe the non-dissipative drag between superfluids~\cite{andreev_three-velocity_1975,fil_nondissipative_2005}.
Within our truncation, the interaction between both species of bosons 
is driven exclusively by the coupling $\lambda_{AB}$, which 
only couples the phase-independent terms $\rho_a$. 
However, because phase fluctuations are not too strong in two and three
dimensions at zero temperature~\cite{andersen_phase_2002}, the superfluid drag
does not considerably affect their macroscopic properties. For details on the effect of the superfluid drag in one dimension see Ref.~\cite{parisi_spin_2018}.

\subsection{Initial conditions and physical inputs}
\label{sec:FRG;sub:IC}

The flow is started at a UV scale $\Lambda$ much larger than the physical scale
set by the chemical potential. We obtain the initial conditions of the RG flow
by imposing that $\Gamma_\Lambda=\mathcal{S}$. For the homogeneous (miscible) 
phase, these are
\begin{equation}
 \begin{gathered}
 S(\Lambda)=Z(\Lambda)=1,\qquad V(\Lambda)=0,\\
 \rho_0(\Lambda)=n_0(\Lambda)=\frac{\mu}{\lambda(\Lambda)+\lambda_{AB}(\Lambda)},\qquad n_1(\Lambda)=1,
  \end{gathered}
\end{equation}
where $\mu>0$. The interaction terms $\lambda$ and $\lambda_{AB}$ 
need to be renormalized so they can be connected to the $s-$wave scattering 
lengths $a$ and $a_{AB}$ associated to the intra- and inter-species 
interactions, respectively. With this, the initial conditions
are completely defined in terms of the physical inputs $\mu$,
$a$ and $a_{AB}$.
We impose that the interaction terms at $k=0$ 
in vacuum are equal to the known two-body $T$-matrices: 
$\lambda(0)=T^{2B}$, $\lambda_{AB}(0)=T_{AB}^{2B}$~\cite{floerchinger_functional_2008}.
In two and three dimensions these are~\cite{stoof_ultracold_2009}
\begin{equation}
    T^{2B}_\alpha=
\begin{cases}
\dfrac{4\pi/m}{\log(-2/m|\mu| a_\alpha^2)-2\gamma_E} & :d=2\\
\dfrac{4\pi a_\alpha}{m} & :d=3
\end{cases},
\label{sec:FRG;sub:flow_eqs;eq:T2B}
\end{equation}
where $T^{2B}_\alpha=T^\text{2B},T^\text{2B}_{AB}$, $a_\alpha=a,a_{AB}$ and
$\gamma_E\approx 0.577$ is the Euler-Mascheroni constant.
For the optimized regulator~(\ref{sec:FRG;sub:flow_eqs;eq:Litim}) we 
obtain the following initial conditions~\cite{floerchinger_functional_2008,isaule_thermodynamics_2020}
\begin{equation}
   \lambda_\alpha(\Lambda)=
\begin{cases}
\dfrac{4\pi/m}{1-2\gamma_E-\log(a_\alpha^2\Lambda^2/4)} & :d=2\\
\dfrac{4\pi/m}{a_\alpha^{-1}-4\Lambda/3\pi} & :d=3
\end{cases}\,,
\end{equation}
where $\lambda_\alpha=\lambda,\lambda_{AB}$. Note that although we have
to choose a high starting scale $\Lambda$, for repulsive interactions
we must choose $\Lambda\lesssim a_\alpha^{-1}$, otherwise
$\lambda_\alpha(\Lambda)<0$ (for details see Ref.~\cite{isaule_thermodynamics_2020}).

Furthermore, we must also choose $a_{AB}<a$ so that the
spin mass term $2\rho_0(\lambda-\lambda_{AB})>0$ at the starting
scale $\Lambda$ (see next subsection). 
A stronger inter-species repulsion corresponds
to a flow that starts in the separable (non-miscible) phase~\cite{pitaevskii_bose-einstein_2016}, which is not described
by our ansatz.
In  subsection~\ref{sec:results;sub:kh} we show that the FRG flow signals
the physical separable phase by the vanishing of the spin healing scale~\cite{tommasini_bogoliubov_2003}.

\subsection{Scale regimes}
\label{sec:FRG;sub:kh}

As the momentum scale $k$ is lowered, the RG flow gradually incorporates into $\Gamma_k$ the fluctuations associated with different scales. Therefore, we can identify different scale
regimes depending on the relevant physics. 

In a one-component Bose gas with broken $U(1)$-symmetry,
the flow can be separated into a Gaussian regime (for $Z k^2/2m \ll 2\lambda\rho_0$)
where both longitudinal and Goldstone fluctuations play a similar role, and
a Goldstone regime (for $Z k^2/2m \ll 2\lambda\rho_0$)
where the flow is dominated by Goldstone fluctuations~(for details see Ref.~\cite{wetterich_functional_2008}).
In order to identify the analogous regimes in a Bose-Bose mixture, we
examine the behavior of the determinant of $\MG^{-1}_k$~(\ref{sec:FRG;sub:ansatz;eq:Ginv}). 
Within our truncation, it reads
\begin{multline}
    \det(\MG_k^{-1})=\left(S^2 \omega^2+(V\omega^2+ E^{(+)}_{1,k})(V\omega^2+ E_{2,k})\right)\\
    \times\left(S^2 \omega^2+(V\omega^2+ E^{(-)}_{1,k})(V\omega^2+ E_{2,k})\right),
    \label{sec:FRG;sub:kh;eq:detGinv}
\end{multline} 
where 
\begin{align}
    E^{(\pm)}_{1,k}(\Q)=&E_{1,k}(\Q)\pm_1 2\lambda_{AB}\rho_0,\nonumber\\
    =&Z\frac{\Q^2}{2m}+2(\lambda\pm\lambda_{AB})\rho_0+R_k(\Q) 
    \label{sec:FRG;sub:kh;eq:E1pm}
\end{align}
and $E_{1,k}$ and $E_{2,k}$ are defined in Eqs.~(\ref{sec:FRG;sub:ansatz;eq:E1}) and (\ref{sec:FRG;sub:ansatz;eq:E2}). Note that we take $\tilde{\mu}=\mu$.

The poles of the propagator and the dispersion relations are extracted from $\det(\MG_k^{-1})=0$
(see App.~\ref{app:Dispersion}). From Eq.~(\ref{sec:FRG;sub:kh;eq:detGinv}) we see that
there are two solution branches. The positive branch ($E^{(+)}_{1,k}$) corresponds to the
density (in-phase) mode, whereas the negative branch ($E^{(-)}_{1,k}$) corresponds to the
spin (out-of-phase) mode~\cite{armaitis_hydrodynamic_2015}.

It is easy to see that the dependency on momentum in Eq.~(\ref{sec:FRG;sub:kh;eq:detGinv})
depends on how the mass-like terms $2(\lambda\pm\lambda_{AB})\rho_0$ compare to the kinetic terms.
Thus, we can identify different scale regimes from the dimensionless quantities
\begin{equation}
    \omega_{h,\pm}=\frac{Z^2 k^2/2m}{2(\lambda\pm\lambda_{AB})\rho_0},
    \label{sec:FRG;sub:kh;eq:omegas}
\end{equation}
which for $\lambda_{AB}=0$ recovers the analogous expression in one-component gases~\cite{wetterich_functional_2008}.
We define the momentum \emph{healing scales} $k_{h,\pm}$ from the points in the flow
where $\omega_{h,\pm}=1$. Note that even though we define $k_{h,\pm}$ in analogy to the physical healing lengths~\cite{pitaevskii_bose-einstein_2016}, $k_{h,\pm}$ correspond to scales in
the RG flow and are not extracted from the physical inputs.

Similarly to one-component gases, the Gaussian regime is 
defined for high scales $k\gg k_{h,\pm}$ ($\omega_{h,\pm}\gg 1$) where the mass-like terms $2(\lambda\pm\lambda_{AB})\rho_0$ are small and
$E^{(\pm)}_{1,k}\approx E_{2,k}$. At these high scales many-body effects are not important,
and thus the flows are similar to those in vacuum~\cite{floerchinger_functional_2008}.
On the opposite side, the
Goldstone regime is defined for low scales $k\ll k_{h,\pm}$ ($\omega_{h,\pm}\ll 1$) where
$E^{(\pm)}_{1,k}\approx 2(\lambda\pm\lambda_{AB})\rho_0$ and
Goldstone fluctuations become dominant.

Between these two regimes, at scales $k_{h,-}\lesssim k \lesssim k_{h,+}$, 
the mixture develops an additional regime where  the effect of the interaction 
between both species of bosons becomes most important.
In particular, around $k_{h,+}$ and $k_{h,-}$ fluctuations of the density and spin modes dominate, respectively. This is easy to see in the cases where $k_{h,-}\ll k_{h,+}$.
Here, around $k_{h,+}$ the spin mode is in a Gaussian-like regime as
$E^{(-)}_{1,k}\approx E_{2,k}$, whereas the density mass term $2(\lambda+\lambda_{AB})\rho_0$
is of the order of the kinetic terms and density fluctuations become important. 
On the hand, around $k_{h,-}$ the density mode is in a Goldstone-like regime $E^{(+)}_{1,k}\approx 2(\lambda+\lambda_{AB})\rho_0$ and spin fluctuations dominate.

We provide additional discussion in App.~\ref{app:Dispersion} where we examine the dispersion relations extracted from Eq.~(\ref{sec:FRG;sub:kh;eq:detGinv}).

\section{Results}
\label{sec:results}

In this section, we present flows of the interaction terms $\lambda$ and $\lambda_{AB}$. We analyze
the scale regimes and the phase separation point. 
We also present results for macroscopic thermodynamic properties
and compare with analytical results from perturbative approaches.

\subsection{Flow of the interactions}
\label{sec:results;sub:flows}

The RG flows of the mixture have similar properties to those of one-component gases~\cite{dupuis_non-perturbative_2007,floerchinger_functional_2008},
Therefore, here we focus on the new features. Details of the RG flows are given in App.~\ref{app:Flows}.

The intra-species interaction coupling $\lambda$ maintains its one-component behavior for small $k$ in
the Goldstone regime: $\lambda$ vanishes for $k\to 0$ linearly with $k$ in two dimensions and logarithmically in three dimensions~\cite{dupuis_non-perturbative_2007}.
This is required to satisfy the vanishing of the anomalous self-energy~\cite{nepomnyashchii_contribution_1975,dupuis_infrared_2011}. 
Similarly, we observe that $\lambda_{AB}$ vanishes for $k\to 0$ as well. However, it is relevant to examine how its flow compares to that of $\lambda$.

\begin{figure*}[t!]
\centering
 \subfloat[Two dimensions]{\includegraphics[scale=0.65]{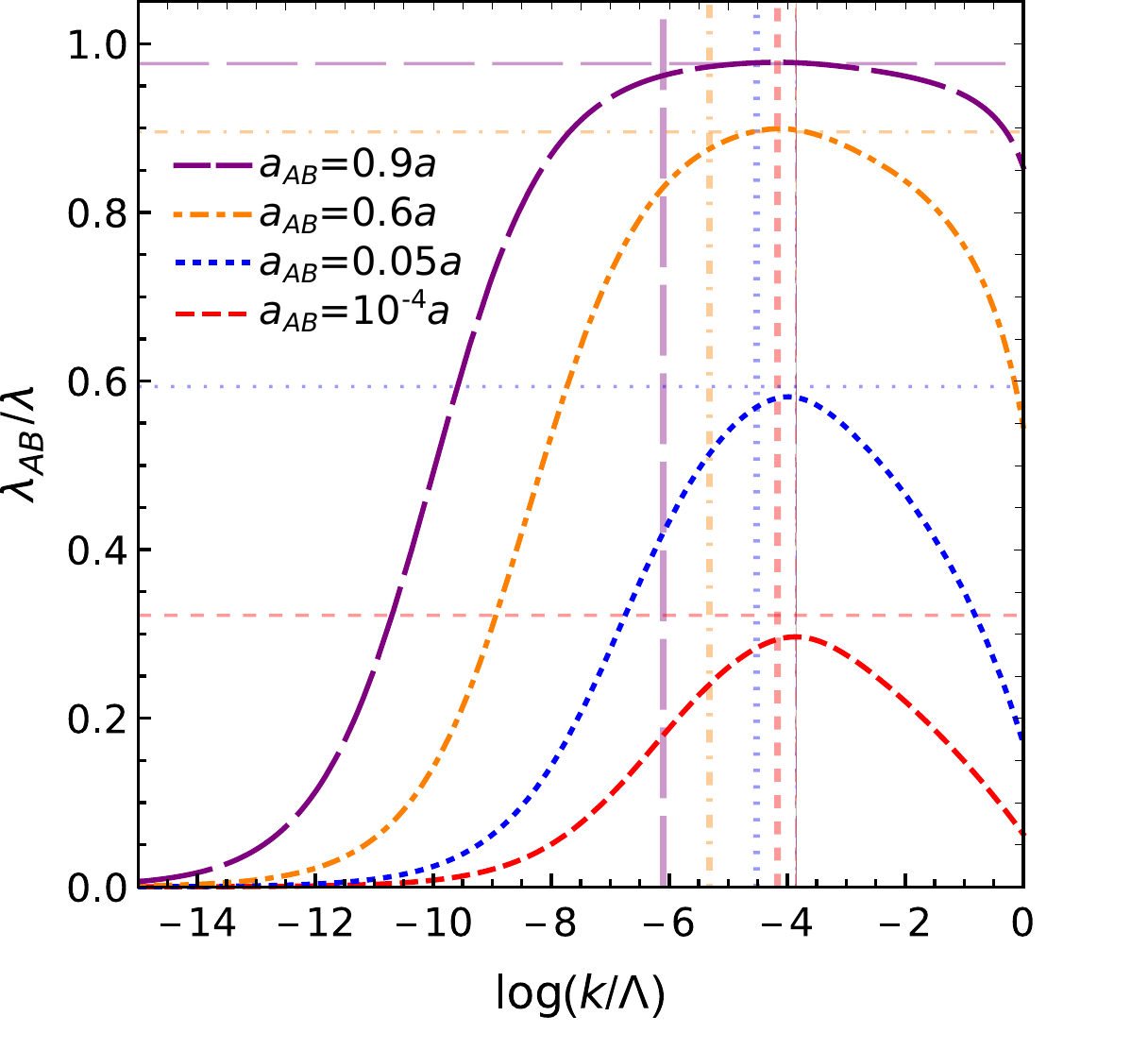}}
 \subfloat[Three dimensions]{\includegraphics[scale=0.65]{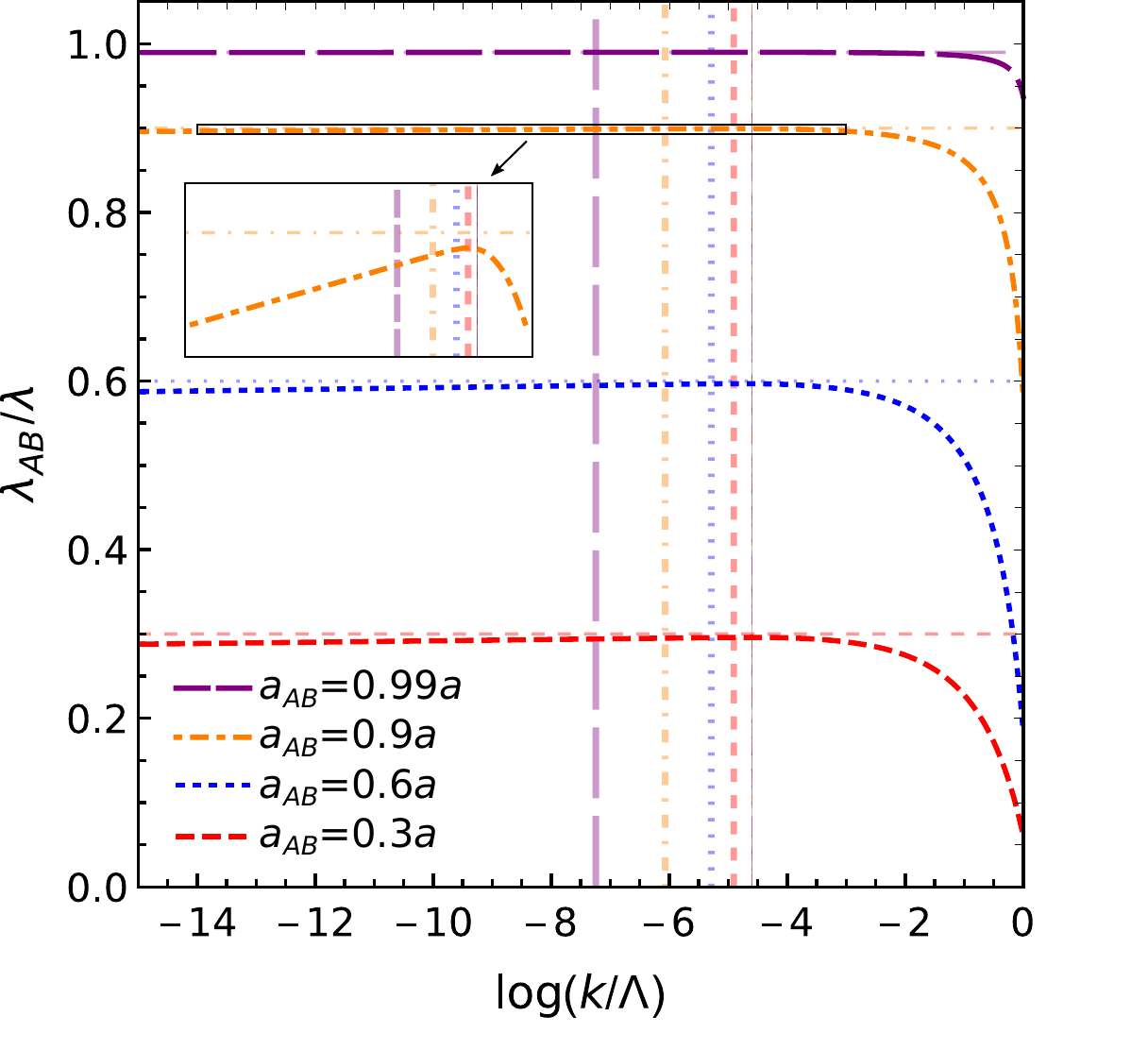}}
\caption{Flows of the ratio $\lambda_{AB}/\lambda$ for different values of
$a_{AB}/a$ as a function of $k$ in two
and three dimensions for $m a^2 \mu=10^{-4}$. 
The four thick vertical lines on the left denote $k_{h,-}$, whereas the 
thin vertical lines on the right (on top of each other) denote $k_{h,+}$.
The horizontal lines correspond to the ratios
$T^{2B}_{AB}/T^{2B}$, where these are defined in Eq.~(\ref{sec:FRG;sub:flow_eqs;eq:T2B}).
The inset in (b) enlarges one of the curves to show its logarithmic flow. The other
curves show similar logarithmic behavior.
\label{sec:results;sub:flows;fig:lABol}}
\end{figure*}

Fig.~\ref{sec:results;sub:flows;fig:lABol} shows examples of flows of the ratio
$\lambda_{AB}/\lambda$ in two and three dimensions for different values of $a_{AB}/a$
(for the individual flows see Fig.~\ref{app:Flows;sub:Flows;fig:lambdas}).
Note that we use a different range of $a_{AB}/a$ in two and three dimensions
because of the different dependence on the scattering length (see Eq.~(\ref{sec:FRG;sub:flow_eqs;eq:T2B})).
We observe that this ratio shows distinct features during the different regimes of the flow. 
At high scales the ratio increases as we lower $k$ until it reaches
its MF value $\approx T^{2B}/T_{AB}^{2B}$ around $k_{h,+}$, showing
the dependence on $a_{AB}$.
On the other hand, for small scales $k \lesssim k_{h,-}$ the ratio decreases
until it vanishes for $k\to 0$. 
We obtain that $\lambda_{AB}/\lambda$ vanishes for small scales linearly with $k$
in two dimensions and logarithmically in three dimensions.

Because $\lambda_{AB}/\lambda$ vanishes for small $k$, 
the density and spin branches in Eq.~(\ref{sec:FRG;sub:kh;eq:detGinv}) are
not separated for $k\to 0$ and both components of the mixture decouple
at long distances. This may seem counterintuitive from a Bogoliubov perspective,
where both modes are always present in a perturbative expansion. 
However, in our RG formalism, the fluctuations of density and spin modes are
incorporated around their associated scales $k_{h,+}$ and $k_{h,-}$,
and the effective action at long distances do not necessarily maintain
the form of the microscopic theory.

It is also worth noting the behavior of $k_{h,\pm}$. The density healing scale $k_{h,+}$ 
is roughly independent of $a_{AB}$ for a fixed chemical potential. On the other hand,
the spin healing scale $k_{h,-}$ depends strongly on the inter-species interaction, decreasing
as $a_{AB}$ increases. Furthermore, we observe that $\lambda_{AB}/\lambda\approx T^{2B}/T_{AB}^{2B}$ for a larger range of scales as $k_{h,-}$ decreases, approaching the solution $\lambda_{AB}/\lambda\to 1$ for $k\to 0$ at the phase separation point $a_{AB}=a$. 
We examine this in the following.

\subsection{Healing scales and phase separation}
\label{sec:results;sub:kh}

The condition for phase-separation during the RG flow corresponds to 
$\lambda=\lambda_{AB}$~\cite{kolezhuk_stability_2010}. 
Under this condition, from Eq.~(\ref{sec:FRG;sub:kh;eq:omegas}) we have that
$\omega_-\to\infty$, and thus the spin healing scale vanishes $k_{h,-}\to 0$.
In contrast, the density healing scale remains finite. 
The same condition applies within perturbative approaches, where the similarly
defined momentum healing scales are defined by~\cite{oles_n_2008,chiquillo_equation_2018}
\begin{equation}
    p^2_{h,\pm}=4m\mu \frac{g\pm g_{AB}}{g+g_{AB}}.
    \label{sec:results;sub:kh;eq:kh_MF}
\end{equation}
At the MF phase separation point $g_{AB}=g$, we have $p_{h,-}=0$ and $p_{h,+}>0$.
Therefore, to study the phase-separation point, we can examine the behavior of $k_{h,-}$ for different physical inputs.

\begin{figure*}[t!]
\centering
 \subfloat[Two dimensions]{\includegraphics[scale=0.65]{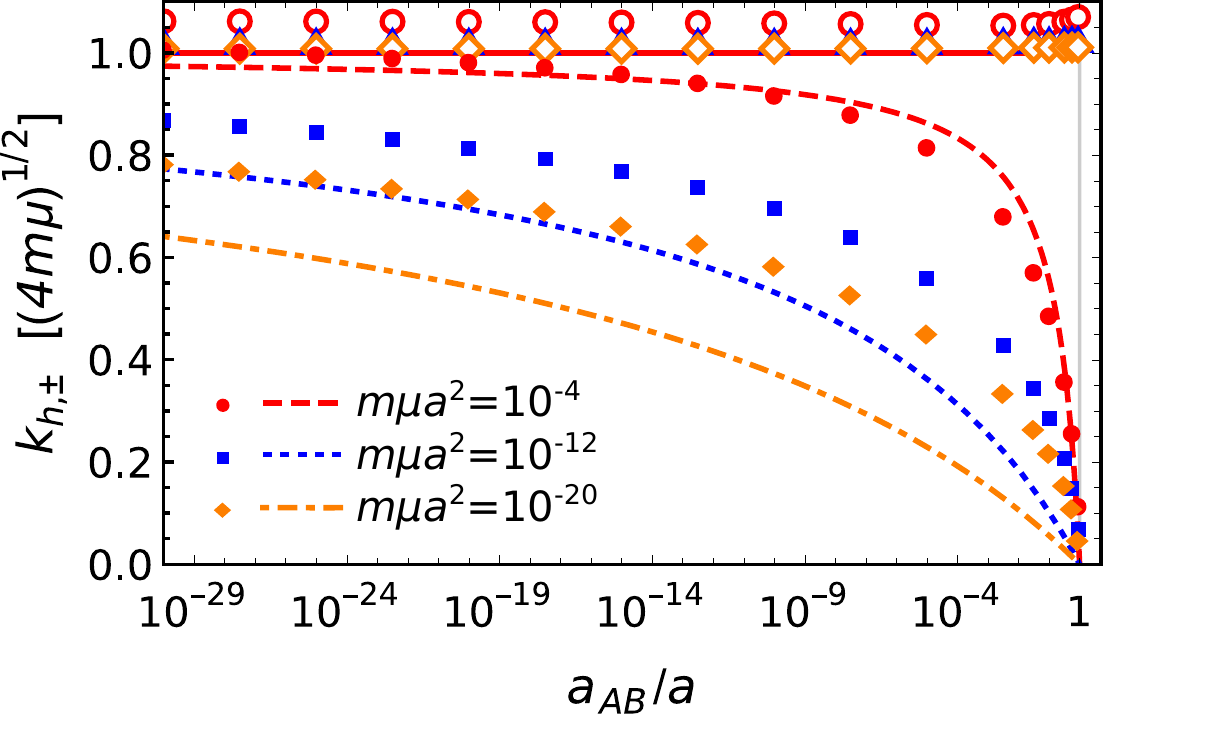}}
 \subfloat[Three dimensions]{\includegraphics[scale=0.65]{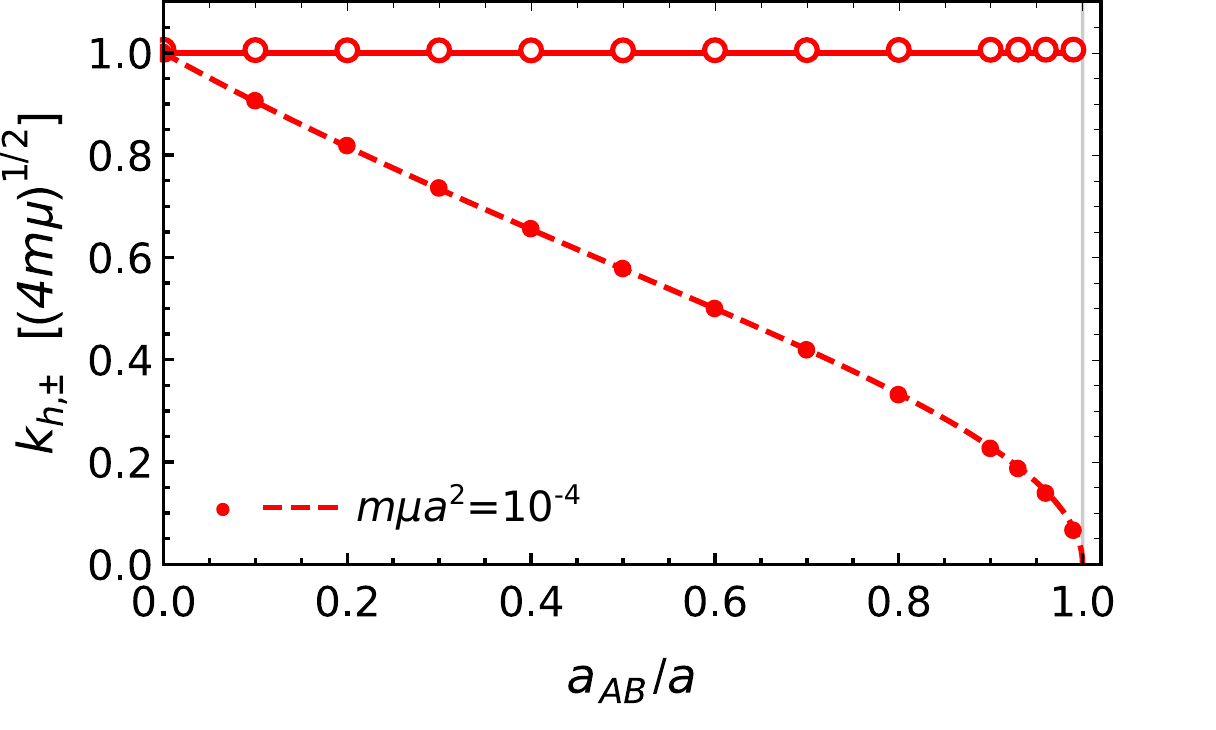}}
\caption{Healing scales $k_{h,\pm}$ in two and three dimensions as a function of
$a_{AB}/a$. Circles correspond to FRG results for $k_{h,+}$ (open) and $k_{h,-}$ (filled), whereas
lines correspond to the Bogoliubov healing scales $p_{h,\pm}$~(\ref{sec:results;sub:kh;eq:kh_MF}). 
In three dimensions the healing scale is independent of the chemical
potential, and thus we plot one choice of $m\mu a^2$.
\label{sec:results;sub:kh;fig:kh}}
\end{figure*}

Fig.~\ref{sec:results;sub:kh;fig:kh} shows healing scales for a range
of inter-species scattering lengths $a_{AB}$. We compare our FRG
calculations with the Bogoliubov healing scales (\ref{sec:results;sub:kh;eq:kh_MF}),
where we use the $T$-matrices defined in Eq.~(\ref{sec:FRG;sub:flow_eqs;eq:T2B}) as
the interactions $g$ and $g_{AB}$. 

As discussed in the last subsection, the density healing scale $k_{h,+}$
is roughly independent of $a_{AB}/a$, as expected for the balanced mixture. On the other hand, the spin healing scale $k_{h,-}$ decreases as $a_{AB}$ increases.
We obtain  that $k_{h,-}$ vanishes for $a_{AB}\to a$.
Therefore, we obtain that the phase separation occurs for equal inter- and intra-species interactions
$a_{AB}=a$, coinciding with the MF prediction. 
We can understand this from the behavior of $\lambda_{AB}/\lambda$ discussed in the last subsection.
Because this ratio decreases for $k<k_{h,-}$, then the condition $\lambda_{AB}=\lambda$
is satisfied only for $a_{AB}=a$.

Our result contrasts with the phase-separation point obtained with RG 
methods in Ref.~\cite{kolezhuk_stability_2010},
where it is stated that in two dimensions, the phase separation occurs for $a_{AB}<a$ at logarithmically low densities.
However, that work study the mixture around the zero-density critical point, and thus
we might not be exploring that regime.
An FRG study of the quantum phase transition of the mixture around the zero-density point
can be done by extending the same analysis for one-component gases from Ref.~\cite{wetterich_functional_2008}.
However, this is beyond the scope of the current work.

\subsection{Thermodynamics}
\label{sec:results;sub:thermo}

In the following we present results for the energy per particle $E/N$
and the condensation depletion $\Delta\Omega_c=1-\Omega_c$.
$E/N$ is computed from the physical values of the pressure, density and chemical
potential as follows from  Eq.~(\ref{sec:S;eq:E}).
For details on how to extract the pressure see Ref.~\cite{isaule_thermodynamics_2020}.

\subsubsection{Three dimensions}

\begin{figure}[t!]
\centering
\includegraphics[scale=0.65]{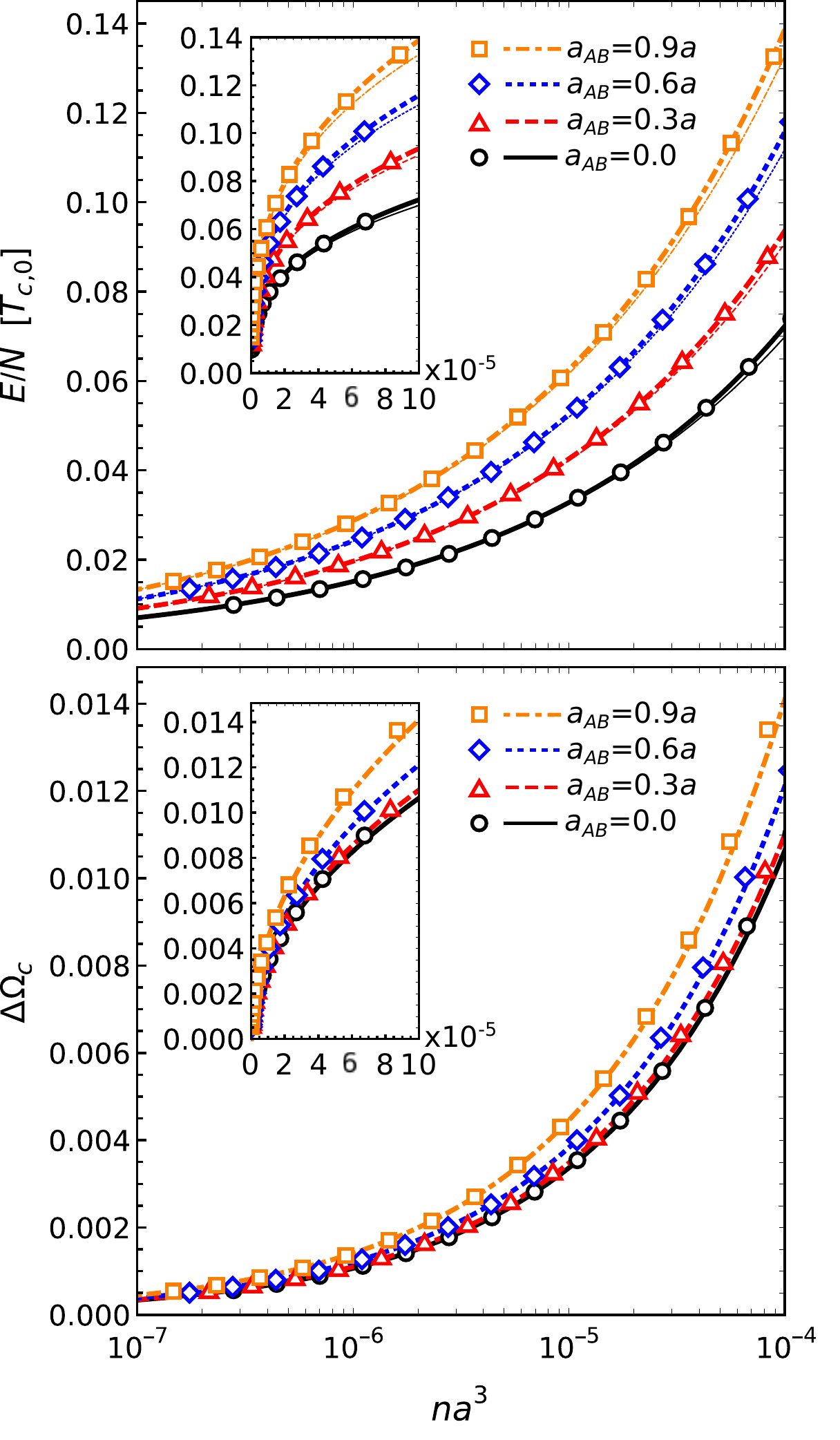}
\caption{Energy per particle $E/N$ and condensate depletion $\Delta\Omega_c$ in three dimensions 
as a function of the concentration parameter $n a^3$, where
$n=2n_0$ is the total density of atoms. 
$E/N$ is scaled in terms of the critical temperature of an
ideal Bose gas $T_{c,0}=\frac{2\pi}{m}(n_0/\zeta(3/2))^{2/3}$.
Markers correspond to FRG results, 
thin lines to the MF solutions and thick lines to MF+LHY solutions~(\ref{sec:results;sub:thermo;eq:EoN3D}) and (\ref{sec:results;sub:thermo;eq:DOc}). 
The insets show the results in
a linear scale.}
\label{sec:results;sub:thermo;fig:thermo3D}
\end{figure}

Fig.~\ref{sec:results;sub:thermo;fig:thermo3D} shows results for $E/N$ and $\Delta\Omega$
in three dimensions for a range of  intra-species scattering lengths $a$ for different ratios of $a_{AB}/a$. Note that $a_{AB}=0$ corresponds
to the one-component limit.
We compare the energy per particle with the perturbative result~\cite{larsen_binary_1963}
\begin{equation}
   \frac{E}{N}=\frac{\pi n}{m}\left(a+a_{AB}\right)+\frac{32\sqrt{2\pi}}{15}\frac{n^{3/2}a^{5/2}}{m}f(a_{AB}/a),\label{sec:results;sub:thermo;eq:EoN3D}   
\end{equation}
where $f(x)=(1+x)^{5/2}+(1-x)^{5/2}$ and $n=2n_0$ is the total density.
The first term corresponds to the MF solution and the second term
to the LHY correction. Similarly, we compare the condensate depletion with the LHY
result~\cite{ota_thermodynamics_2020}
\begin{equation}
    \Delta\Omega_c = \frac{4}{3\sqrt{2\pi}}(na^2)^{3/2}h(a_{AB}/a),
    \label{sec:results;sub:thermo;eq:DOc}   
\end{equation}
where $h(x)=(1+x)^{3/2}+(1-x)^{3/2}$.

We obtain an excellent agreement with the LHY results, 
showing that our calculations correctly incorporate the effect of quantum fluctuations and of the inter-species interaction.
We stress that the LHY corrections give an accurate description of
Bose gases in three dimensions for this range of $na^3$~\cite{andersen_theory_2004}.
In particular, the accuracy of the LHY result for the condensate depletion in
one-component gases has been proved experimentally~\cite{lopes_quantum_2017}.

\subsubsection{Two dimensions}

\begin{figure}[t!]
\centering
\includegraphics[scale=0.65]{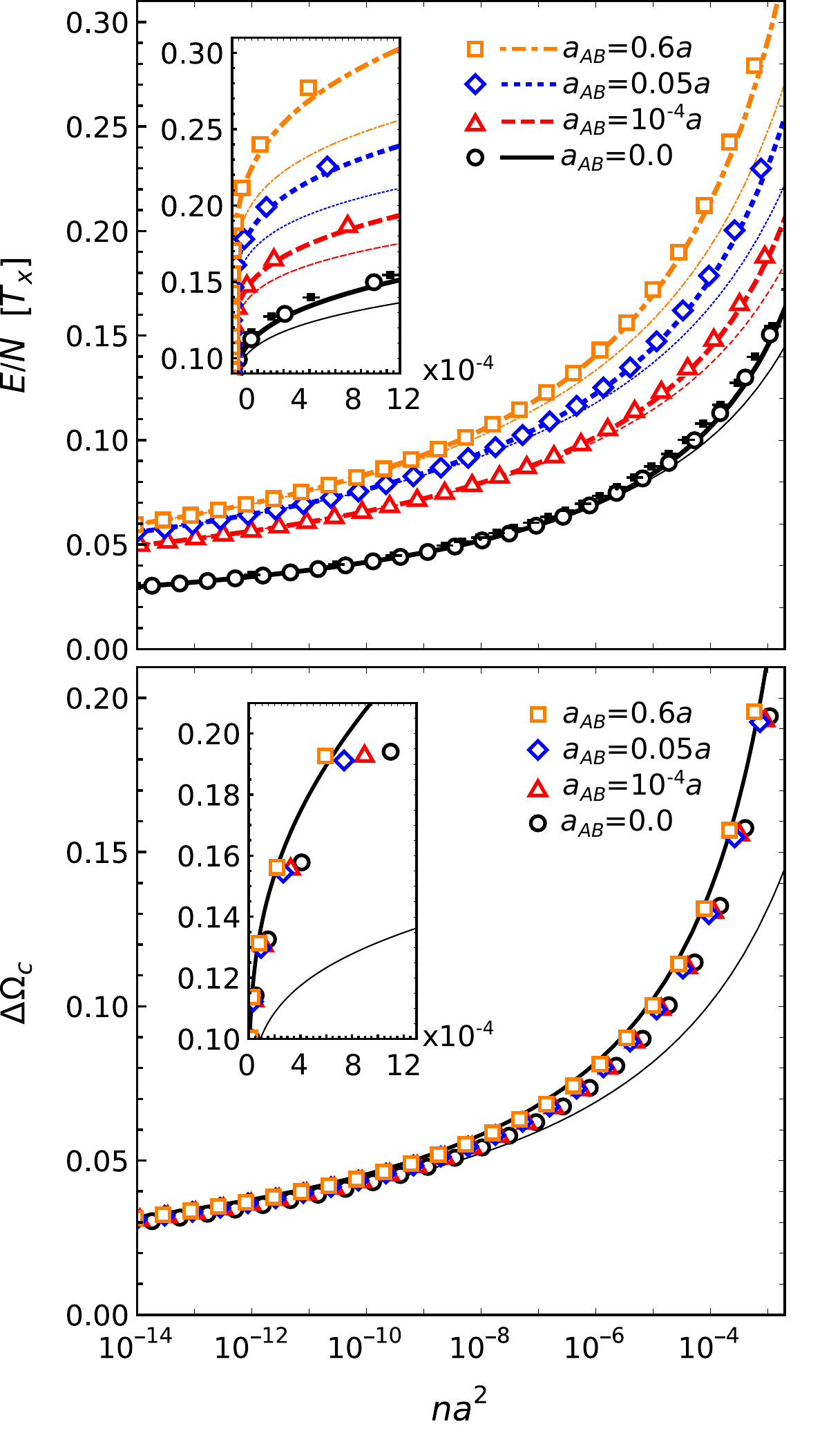}
\caption{Energy per particle $E/N$ and condensate depletion $\Delta\Omega_c$ in two dimensions 
as a function of the concentration parameter $n a^2$, where
$n=2n_0$ is the total density of atoms. 
$E/N$ is scaled in terms of the characteristic temperature $T_x=2\pi n_0/m$.
Markers correspond to FRG results. (Top):
Thin lines correspond to the MF solution and thick lines to MF+LHY solution~(\ref{sec:thermo;sub:2D;eq:EoNLHY}), whereas the filled black
circles correspond to MC simulations for the one-component
gas from Ref.~\cite{astrakharchik_equation_2009}.
(Bottom): Thin line corresponds to the analytical expression at leading order
$\Delta\Omega_c=1/|\log(na^2/2)|$ and the thick line to Eq.~(\ref{sec:thermo;sub:2D;eq:DOc_2019}).
The insets show the results in
a linear scale.}
\label{sec:results;sub:thermo;fig:thermo2D}
\end{figure}

Fig.~\ref{sec:results;sub:thermo;fig:thermo2D} shows results for $E/N$ and $\Delta\Omega_c$
in two dimension.
We compare the energy per particle with the perturbative result~\cite{konietin_2d_2018}
\begin{multline}
    \frac{E}{N}=\frac{\pi n}{m}\zeta_+(n,a,a_{AB})+\frac{\pi n}{2m}\sum_\pm \zeta^2_\pm(n,a,a_{AB})\\
    \times(2\gamma_E+1/2+\log\left(\pi \zeta_\pm(n,a,a_{AB})\right)
    \label{sec:thermo;sub:2D;eq:EoNLHY}
\end{multline}
where
\begin{equation}
\zeta_\pm(n,a,a_{AB})=\frac{1}{|\log(na^2/2)|}\pm\frac{\Theta(a_{AB})}{|\log(na^2_{AB}/2)|}
\end{equation}
As in three dimensions, the first term in Eq.~(\ref{sec:thermo;sub:2D;eq:EoNLHY}) corresponds
to the MF result and the second to the LHY-type correction. Additionally, we compare
with MC results of $E/N$ for the one-component gas~\cite{astrakharchik_equation_2009}. 
For the condensate depletion, we compare with the recently obtained expression for
the one-component gas~\cite{pastukhov_ground-state_2019}
\begin{equation}
    \Delta\Omega_c =\frac{1}{|\log(na^2/2)|+\log(|\log(na^2/2)|)+C},
    \label{sec:thermo;sub:2D;eq:DOc_2019}
\end{equation}
where $C\approx-\log(\pi)-2\gamma_E-2.63$. 

For $E/N$, we obtain a reasonable agreement between our FRG results and the analytical expression.
We stress that because of the enhanced effect of fluctuations in two dimensions, the LHY-type correction
is not as accurate, and thus we expect that our results give a better description of this system. In particular, for the one-component limit ($a_{AB}=0$), we obtain a slighter better agreement with the MC results, which we can consider exact.
Also, note the larger effect of the LHY correction to the MF result compared to three dimensions, showing
the enhanced effect of quantum fluctuations.

For $\Omega_c$ we obtain that in the one-component limit, our results are comparable to those of
Eq.~(\ref{sec:thermo;sub:2D;eq:DOc_2019}). However, we stress that FRG calculations with better truncations
are needed in order to prove the robustness of our results. Most interesting is the effect of the
inter-species interaction. At first glance it seems that the condensate depletion in two dimensions
is not as sensitive to the inter-species interaction as in three dimensions. However, in our figures, $\Omega_c$
is one order of magnitude greater in two dimensions.
The larger condensate depletion is caused by
the enhanced fluctuations in two dimensions, which are at the critical dimension of destroying the 
condensate~\cite{mermin_absence_1966}. Therefore, in two dimensions, the depletion is mostly
driven by the quantum fluctuations associated with the lower dimensionality, and the effects of the mixture become less noticeably.

\section{Conclusions}
\label{sec:conclusions}

In this work, we study two- and three-dimensional balanced Bose-Bose mixtures at zero temperature within
the FRG. We generalize previous work on one-component gases to now consider two species of bosons
interacting through weak and repulsive intra- and inter-species interactions.

We find that the scale-dependent inter-species interaction coupling vanishes at long distances.
However, the RG flow correctly takes the inter-species interaction effect into account as fluctuations of density and spin modes are incorporated at their associated momentum scales.
We also study the phase separation condition, which in our formalism can be identified by
the vanishing of the momentum spin healing scale. We find that in both two and three dimensions,
the phase separation occurs at the MF point of equal scattering lengths $a_{AB}=a$.
Finally, in order to examine macroscopic thermodynamic properties, we calculate
the energy per particle and condensation depletion for a range of interaction parameters.
We obtain good agreement with analytical results from perturbative approaches. 
We find some deviations from these results in two dimensions, as expected from the enhanced
effect of fluctuations.

Having demonstrated that the FRG is capable of studying Bose-Bose mixture,
we intend to explore in future work how to implement the interaction between the relative phases
of the condensates within our framework. This is necessary to study the mixture at finite temperatures
and the superfluid phase transition where the superfluid drag becomes important~\cite{nespolo_andreevbashkin_2017,karle_coupled_2019}. 
We also intend to study the mixture around the quantum transition at zero density
to check if the MF phase separation point changes as suggested in other works.
Other relevant related extensions are studies of quantum mixtures in different configurations.
Particularly interesting are Bose-Bose mixture with attractive inter-species interaction.
These would enable us to explore the strongly-interacting regime, which presents rich physics,
including droplet phases and dimerization~\cite{hu_microscopic_2020-1,hu_microscopic_2020}. Because perturbative approaches are not suitable
to study this regime, an FRG study could provide a more robust picture of these systems.

\begin{acknowledgments}
This work has been partially supported by MINECO (Spain) Grant No.FIS2017-87534-P. We acknowledge financial support from Secretaria d’Universitats i Recercadel  Departament  d’Empresa  i  Coneixement  de  la  Generalitat  de  Catalunya,  co-funded  by the European Union Regional Development Fund within the ERDF Operational Program of Catalunya (project QuantumCat, ref.  001-P-001644). We acknowledge helpful discussions with Michael C. Birse.
\end{acknowledgments}

\appendix

\section{Flow equations}
\label{app:flow_eqs}

The flow of each coupling is driven by the appropriate projection of the 
Wetterich equation~(\ref{sec:FRG;eq:Wetterich}) into the ansatz~(\ref{sec:FRG;sub:ansatz;eq:Gamma}). 
The flow of the order parameter is driven by the equilibrium condition $\delta\Gamma/\delta\psi_{a,1}=0$, thus
\begin{equation}
    -\sqrt{2\rho_0}\left(\lambda+\lambda_{AB}\right)\dot{\rho}_{0}
 =\frac{\delta\dot{\Gamma}}{\delta \psi_{A,1}}\bigg|_{\rho_0,\mu},
\end{equation}
where $\dot{f}=\partial_k f$. The right-hand-side (RHS) is obtained from taking
the functional derivative to the RHS of the Wetterich equation~(\ref{sec:FRG;eq:Wetterich})
and then evaluating at $\psi_{a,1}(q)=(2\pi)^{d+1}(2\rho_0)^{1/2}\delta(q)$, $\psi_{a,2}(q)=0$ and $\tilde{\mu}=\mu$. The interaction terms are obtained from the flow of the 
longitudinal masses (see Eq.~(\ref{sec:FRG;sub:ansatz;eq:Ginv}))
\begin{align}
   2\rho_0\dot{\lambda}-(\lambda+\lambda_{AB})\dot{\rho}_0
=&\frac{\delta^2\dot{\Gamma}}{\delta \psi^2_{A,1}}\bigg|_{p=0,\rho_0,\mu}\\
2\rho_0\dot{\lambda}_{AB}=
&\frac{\delta^2\dot{\Gamma}}{\delta \psi_{A,1} \delta \psi_{B,1}}\bigg|_{p=0,\rho_0,\mu}, 
\end{align}
where the external momentum $p=(\nu,\vP)$ of the two-point function $\MGamma^{(2)}$ is evaluated
at zero. The renormalization factors are obtained from momentum derivatives of the
two-point function~\cite{floerchinger_functional_2008}
\begin{align}
  \frac{\dot{Z}}{2m}=&\frac{\partial}{\partial \vP^2}
\left(\frac{\delta^2\dot{\Gamma}}{\delta \psi^2_{A,2}}\right)\bigg|_{p=0,\rho_0,\mu},\\
\dot{S}=&\frac{\partial}{\partial \nu}
\left(\frac{\delta^2\dot{\Gamma}}{\delta \psi_{A,2} \delta \psi_{A,1}}\right)\bigg|_{p=0,\rho_0,\mu},\\
\dot{V}=&\frac{\partial}{\partial \nu^2}\left(\frac{\delta^2\dot{\Gamma}}
{\delta \psi^2_{A,2}}\right)\bigg|_{p=0,\rho_0,\mu}.  
\end{align}
Finally, the flow equations for $n_0$ and $n_1$ are obtained from taking derivatives with
respect to $\mu$~\cite{floerchinger_functional_2008}
\begin{align}
    \dot{n}_0-n_1\dot{\rho}_0=&-\frac{\partial}{\partial\mu}\dot{\Gamma},\\
\sqrt{2\rho_0}\dot{n}_1=&-\frac{\partial}{\partial\mu}
\left(\frac{\delta\dot{\Gamma}}{\delta \psi_{A,1}}\right).
\end{align}
For details on the expressions for the RHS's of the flow equations we refer to Ref.~\cite{delamotte_nonperturbative_2004}.

\section{Dispersion relations and microscopic sound velocity}
\label{app:Dispersion}

From the inverse propagator~(\ref{sec:FRG;sub:ansatz;eq:Ginv}) and by performing a continuation
to real time $\omega\to -iq_0$, we can identify the poles of the propagator from
$\det(\MG_k^{-1})=0$. Within our truncation, the propagator has eight poles, which can be summarized as

\begin{widetext}
\begin{align}
    (q^*_{0,1})^2=&\frac{1}{2V^2}\bigg[S^2+V\left(E^{(+)}_{1,k}(\Q)+E_{2,k}(\Q)\right)    -\sqrt{\left(S^2+V\left(E^{(+)}_{1,k}(\Q)+E_{2,k}(\Q)\right)\right)^2-4V^2E^{(+)}_{1,k}(\Q)E_{2,k}(\Q)}\bigg], \label{app:Dispersion:eq:q01}\\
    (q^*_{0,2})^2=&\frac{1}{2V^2}\bigg[S^2+V\left(E^{(-)}_{1,k}(\Q)+E_{2,k}(\Q)\right)    -\sqrt{\left(S^2+V\left(E^{(-)}_{1,k}(\Q)+E_{2,k}(\Q)\right)\right)^2-4V^2E^{(-)}_{1,k}(\Q)E_{2,k}(\Q)}\bigg],\label{app:Dispersion:eq:q02}\\
    (q^*_{0,3})^2=&\frac{1}{2V^2}\bigg[S^2+V\left(E^{(+)}_{1,k}(\Q)+E_{2,k}(\Q)\right)    +\sqrt{\left(S^2+V\left(E^{(+)}_{1,k}(\Q)+E_{2,k}(\Q)\right)\right)^2-4V^2E^{(+)}_{1,k}(\Q)E_{2,k}(\Q)}\bigg],\label{app:Dispersion:eq:q03}\\
    (q^*_{0,4})^2=&\frac{1}{2V^2}\bigg[S^2+V\left(E^{(-)}_{1,k}(\Q)+E_{2,k}(\Q)\right)    +\sqrt{\left(S^2+V\left(E^{(-)}_{1,k}(\Q)+E_{2,k}(\Q)\right)\right)^2-4V^2E^{(-)}_{1,k}(\Q)E_{2,k}(\Q)}\bigg],\label{app:Dispersion:eq:q04}
\end{align}
\end{widetext}
where $E^{(+)}_{1,k}$ is defined in Eq.~(\ref{sec:FRG;sub:kh;eq:E1pm}) and 
$E_{2,k}$ in Eq.~(\ref{sec:FRG;sub:ansatz;eq:E2}).

In the following, we examine the behavior of the poles in the different regimes, 
complementing the discussion in subsection~\ref{sec:FRG;sub:kh}. 
Later, we examine the microscopic sound velocity extracted from the dispersion relation.

\subsection{Scale regimes}

For high scales $k\gg k_{h,\pm}$,  because $E^{(\pm)}_{1,k}\approx E_{2,k}$ 
the density and spin modes are indistinguishable. Furthermore, 
because at these high scales the renormalization factors remain at their microscopic
values $Z,S\approx 1$ and $V\approx 0$ (see App.~\ref{app:Flows}), the propagator
has a single pole
\begin{equation}
    (q_{0}^*)^2=E^2_{2,k}(\Q),
\end{equation}
which recovers the quadratic spectrum at high momentum as in Bogoliubov theory~\cite{pitaevskii_bose-einstein_2016}.

For intermediate scales $k_{h,-}\lesssim k \lesssim k_{h,-}$, it is easy to
see that the inter-species interaction term $\lambda_{AB}$ becomes important.
If we neglect $V$ at these scales, we get the poles
\begin{align}
    (q_{0,+}^*)^2&=E^{(+)}_{1,k}(\Q)E_{2,k}(\Q)/S^2,\\
    (q_{0,-}^*)^2&=E^{(-)}_{1,k}(\Q)E_{2,k}(\Q)/S^2,
\end{align}
analogous to those in perturbative approaches~\cite{chiquillo_equation_2018}.
In particular, if $k_{h,-}\ll k_{h,+}$, 
around $k_{h,+}$ we have a quadratic spectrum for the spin mode $(q_{0,-}^*)^2\approx E^2_{2,k}(\Q)/S^2$, 
whereas around $k_{h,-}$ we have a linear spectrum for the density mode $(q_{0,+}^*)^2\approx2\rho_0(\lambda+\lambda_{AB})E_{2,k}(\Q)/S^2$.

For small scales $k\ll k_{h,\pm}$, the mass terms dominate
$E^{(+)}_{1,k}$ and the spectrum becomes linear.
Particularly insightful is to examine the physical limit $k\to 0$ of the spectrum 
in terms of the sound velocity. We do this in the following.

\subsection{Microscopic sound velocity}

For $k\to 0 $ the spectrum becomes linear with two microscopic sound velocities
$(q^*_{0,\pm})^2=c_\pm \Q^2$~\cite{armaitis_hydrodynamic_2015}. From
the poles of the sound modes~\cite{floerchinger_functional_2008}, 
Eqs.~(\ref{app:Dispersion:eq:q01}) and (\ref{app:Dispersion:eq:q02}), 
we obtain
\begin{equation}
    c_\pm^2=\left(\dfrac{Z/2m}{V+S^2/(2(\lambda\pm\lambda_{AB})\rho_0)}\right)_{k\to 0},
\end{equation}
which for $\lambda_{AB}=0$ takes the known form for a one-component gas~\cite{dupuis_non-perturbative_2007}.
Because $\lambda_{AB},\lambda \to 0$, both sound velocities are equal $c=c_\pm$,
in contrast to the Bogoliubov spectrum that have distinct density and spin sound velocities.
Furthermore, because both $S$ and $\lambda$ vanish with the same scaling behavior~\cite{dupuis_non-perturbative_2007} (see also App.~\ref{app:Flows}),
we obtain that
\begin{equation}
    c^2=\left(\dfrac{Z}{2mV}\right)_{k\to 0},
\end{equation}
as with one-component gases~\cite{dupuis_infrared_2011}. 

First, we stress that the spectrum
remains linear even though the sound velocities differ from that of
Bogoliubov theory. Second, as discussed in the main text, the fact that we do not obtain
separate density and spin sound velocities results from the form of the effective action 
at long distances.  Density and spin fluctuations are correctly incorporated into the macroscopic properties even though the theory
at long distances becomes independent of $\lambda_{AB}$.

\begin{figure*}[t!]
\centering
 \subfloat[Two dimensions]{\includegraphics[scale=0.65]{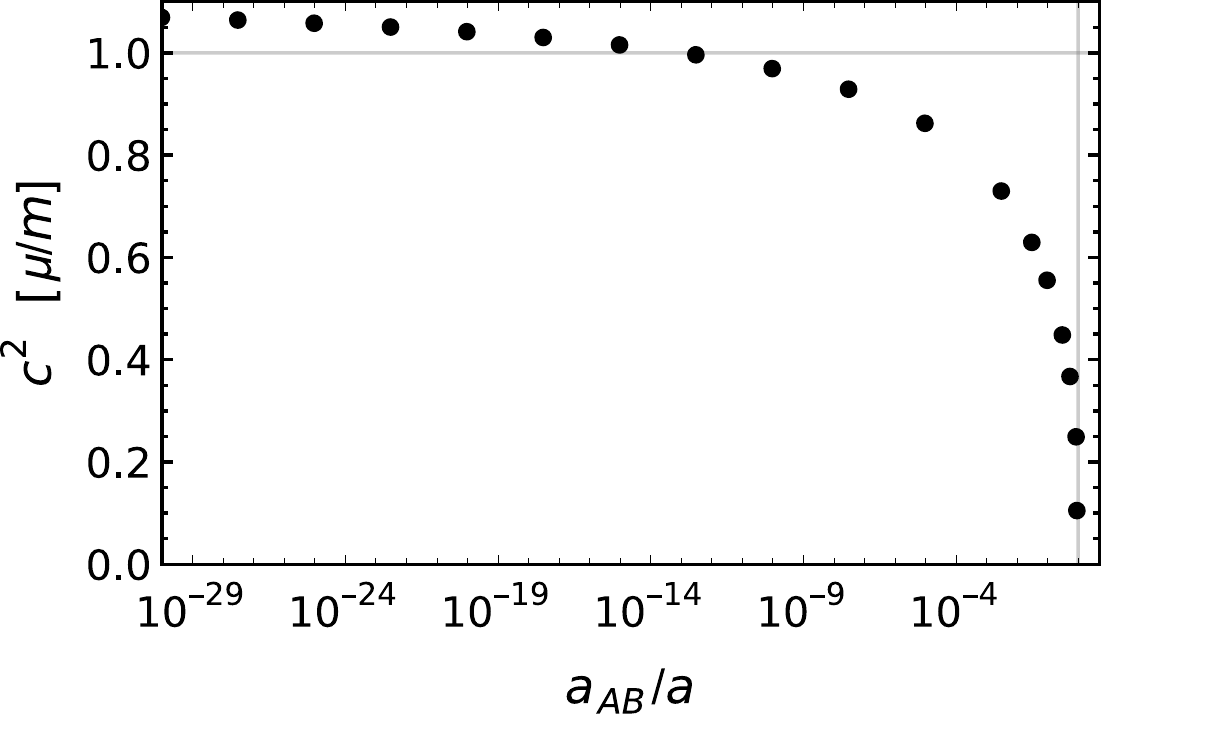}}
 \subfloat[Three dimensions]{\includegraphics[scale=0.65]{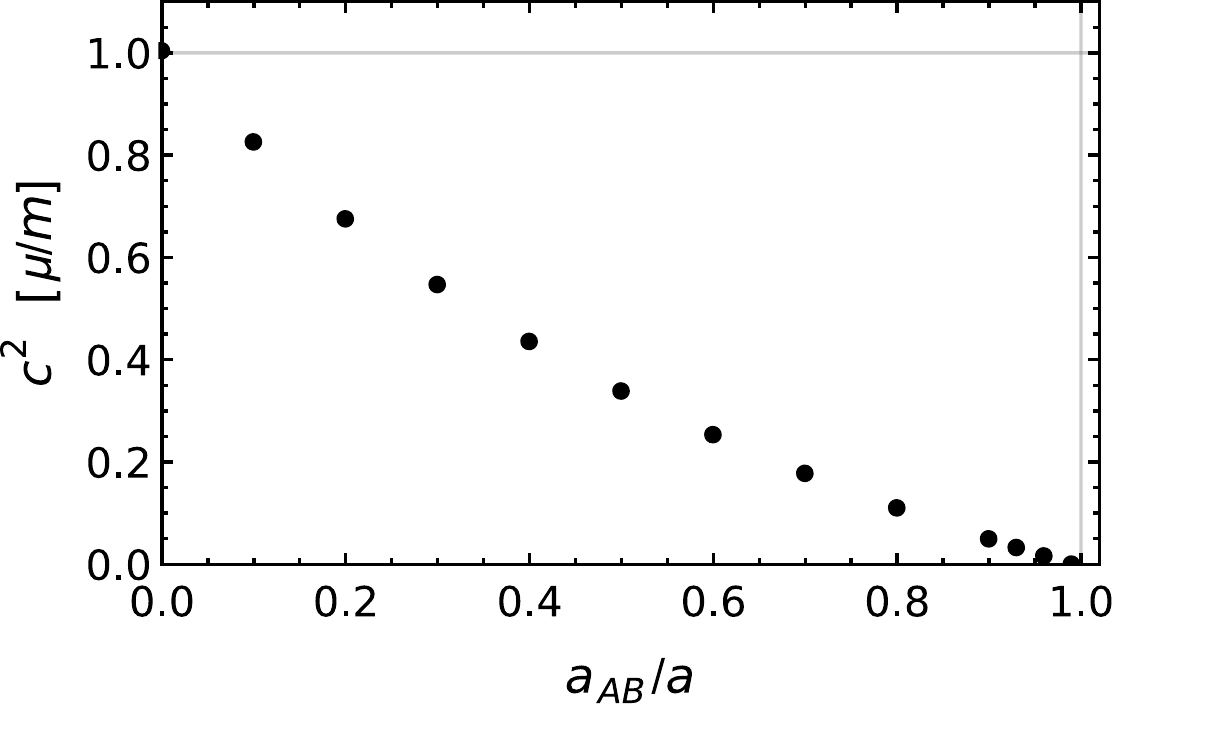}}
\caption{Sound velocity $c^2$ obtained from FRG calculations as a function of $a_{AB}/a$ for $m a^2 \mu=10^{-4}$.
\label{app:Dispersion;fig:k2}}
\end{figure*}

It is also relevant to examine if the sound velocity depends on the inter-species interaction.
In Fig.~\ref{app:Dispersion;fig:k2} we show
$c^2$ in two and three dimensions for a range of $a_{AB}/a$ for a chosen chemical potential.
We obtain that the sound velocity does depend on $a_{AB}$. Moreover, it vanishes for
$a_{AB}\to a$, signaling the phase separation point. It is interesting that,
since the interaction couplings vanish, the decrease
of $c^2$ is driven by the one-component couplings $Z$ and $V$.

Here we note that in a Bogoliubov formulation, the spin sound velocity vanishes at the 
phase separation point, whereas the density sound velocity remains finite~\cite{armaitis_hydrodynamic_2015}. There is no apparent reason beforehand why the FRG flow gives a vanishing microscopic sound velocity at the phase separation.
A detailed analysis of the critical point is required, which is left to future work.

\section{FRG flows}
\label{app:Flows}

As mentioned in subsection~\ref{sec:results;sub:flows}, the flows of the mixture gas are similar to those
of one-component gases (see Refs.~\cite{dupuis_non-perturbative_2007,wetterich_functional_2008,floerchinger_functional_2008} for details). Here we examine the flows of the different $k$-dependent functions. We compare the flows
with calculations using $a_{AB}=0$. In this limit,
$\lambda_{AB}=0$ for \emph{all} $k$ and
the flow equations recover their forms for one-component gases.

\begin{figure*}[t!]
\centering
 \subfloat[Two dimensions]{\includegraphics[scale=0.65]{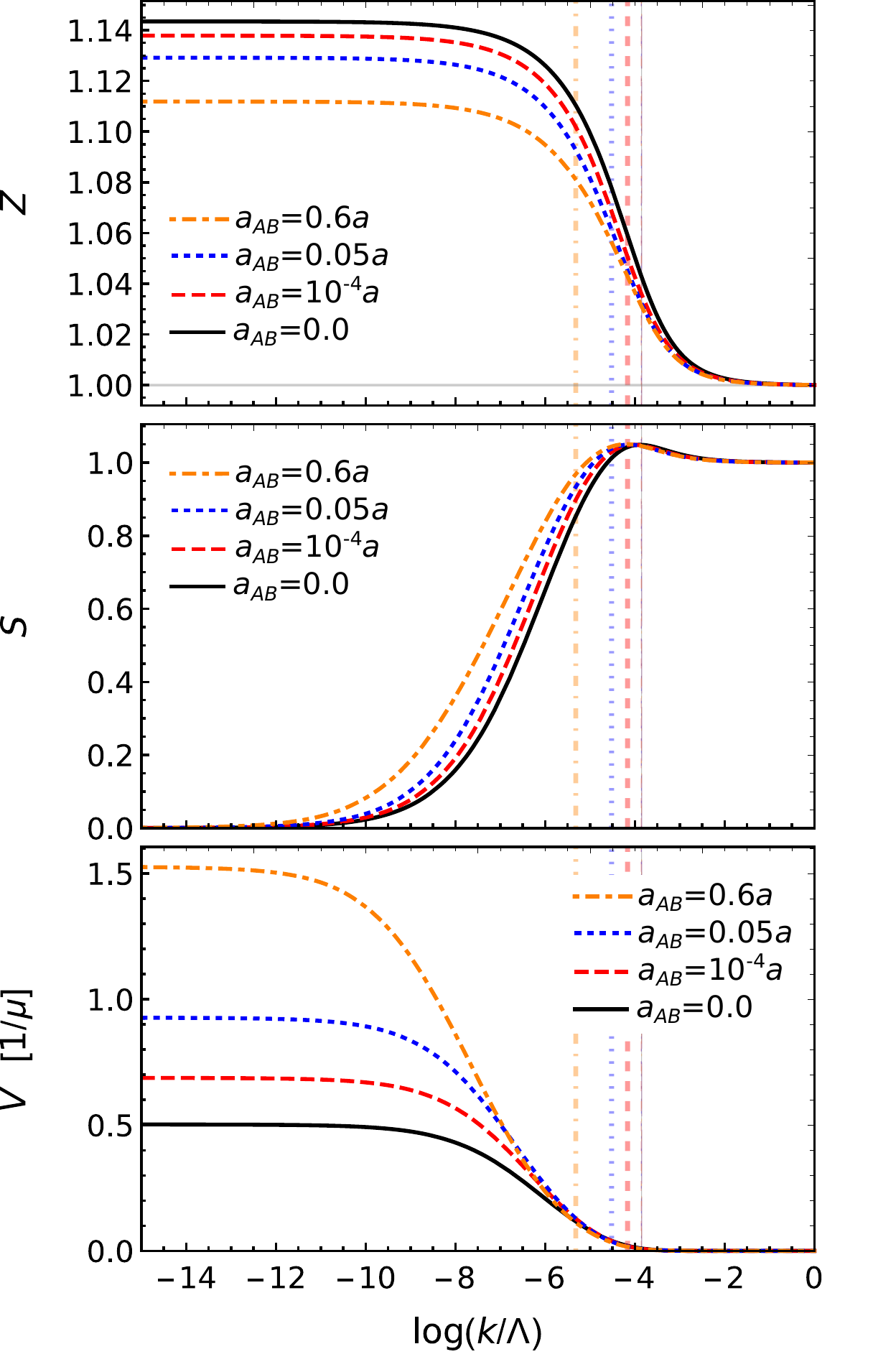}}
 \subfloat[Three dimensions]{\includegraphics[scale=0.65]{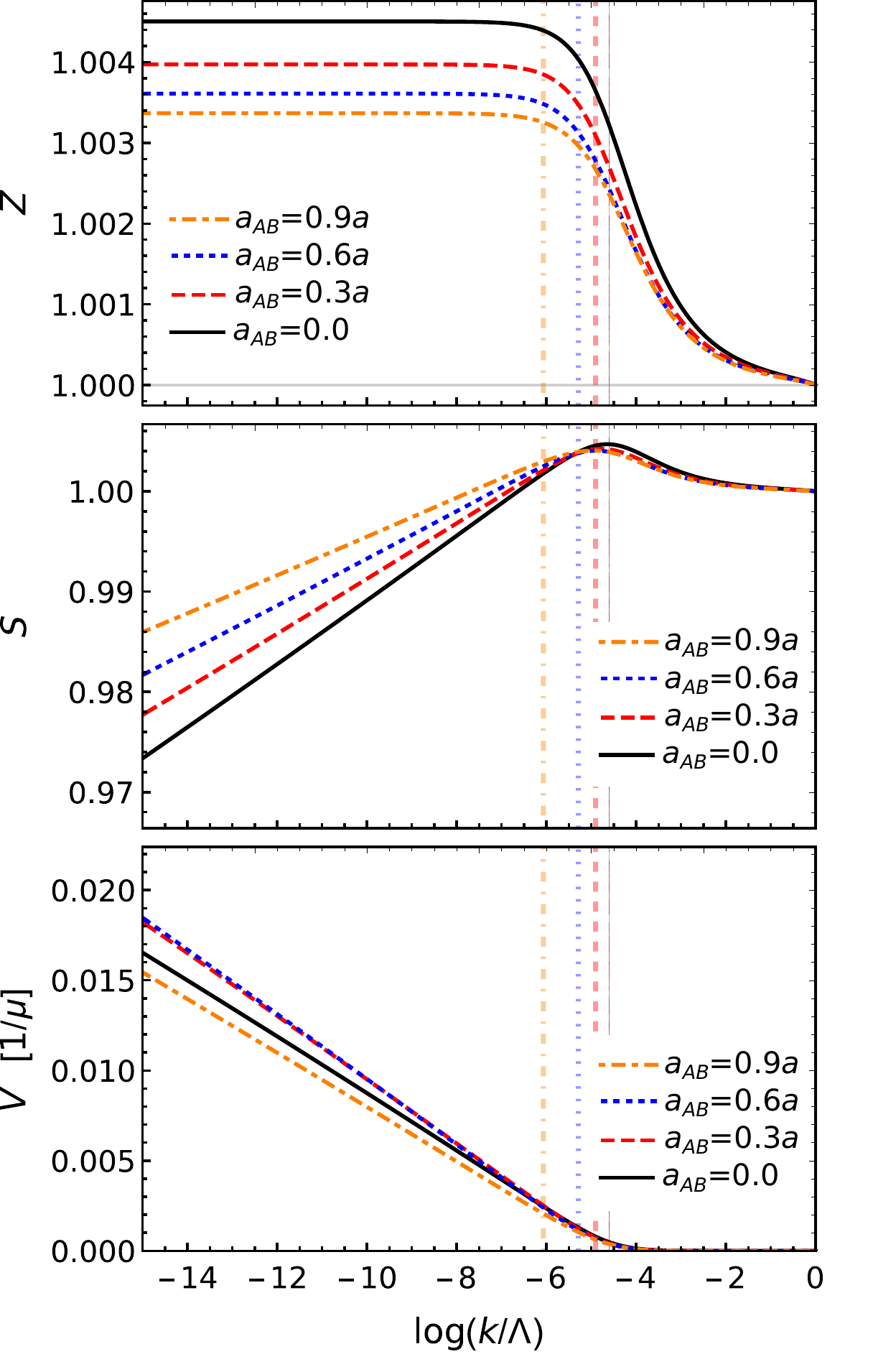}}
\caption{Flows of $Z$, $S$ and $V$ as a function of $k$  in two
and three dimensions for $m a^2 \mu=10^{-4}$. Thin and thick vertical lines denote $k_{h,+}$ and $k_{h,-}$, respectively.}
\label{app:Flows;sub:Flows;fig:kin}
\end{figure*}

Fig.~\ref{app:Flows;sub:Flows;fig:kin} shows flows of the renormalization factors $Z$, $S$
and $V$. The inter-species interaction produces a deviation of all the couplings from
their flows for $a_{AB}=0$. However, the couplings keep their infrared behavior, as discussed in the main text. Also, note that the flows deviate from each
other as $k$ approaches $k_{h,+}$, showing that the inter-species interaction is not important
at higher scales.

The mass renormalization $Z$ increases from its microscopic value at high
scales to converge to a constant for $k\to 0$, whereas the wave-function renormalization $S$ vanishes in the IR 
as $k$ in two dimensions and logarithmically in three dimensions.
The coupling $V$, which is not present in the microscopic action (\ref{sec:S;eq:S}), is generated by the quantum fluctuations as we lower
$k$.
As examined in detail in Ref.~\cite{wetterich_functional_2008}, 
the long distance behavior of the
propagator is dominated by the quadratic frequency term $V$ instead of the linear term
$S$. Thus, the IR is described by a relativistic-like model.
In two dimensions,  it is easy to see that $V$ dominates the IR physics, as $V$ quickly
converges to finite values for $k<k_{h,-}$, whereas $S$ vanishes.
This shows the importance of fluctuations in two dimensions, as the long-distance theory deviates considerably from the Bogoliubov picture.
On the other hand, in three dimensions, the logarithmic flows mean that the coupling $V$ is important only in the extreme IR. Its inclusion does not considerably affect the macroscopic properties. Still, $V$ should converge to a finite value at $k=0$, and its inclusion is important to obtain a consistent theory.

\begin{figure*}[t!]
\centering
 \subfloat[Two dimensions]{\includegraphics[scale=0.65]{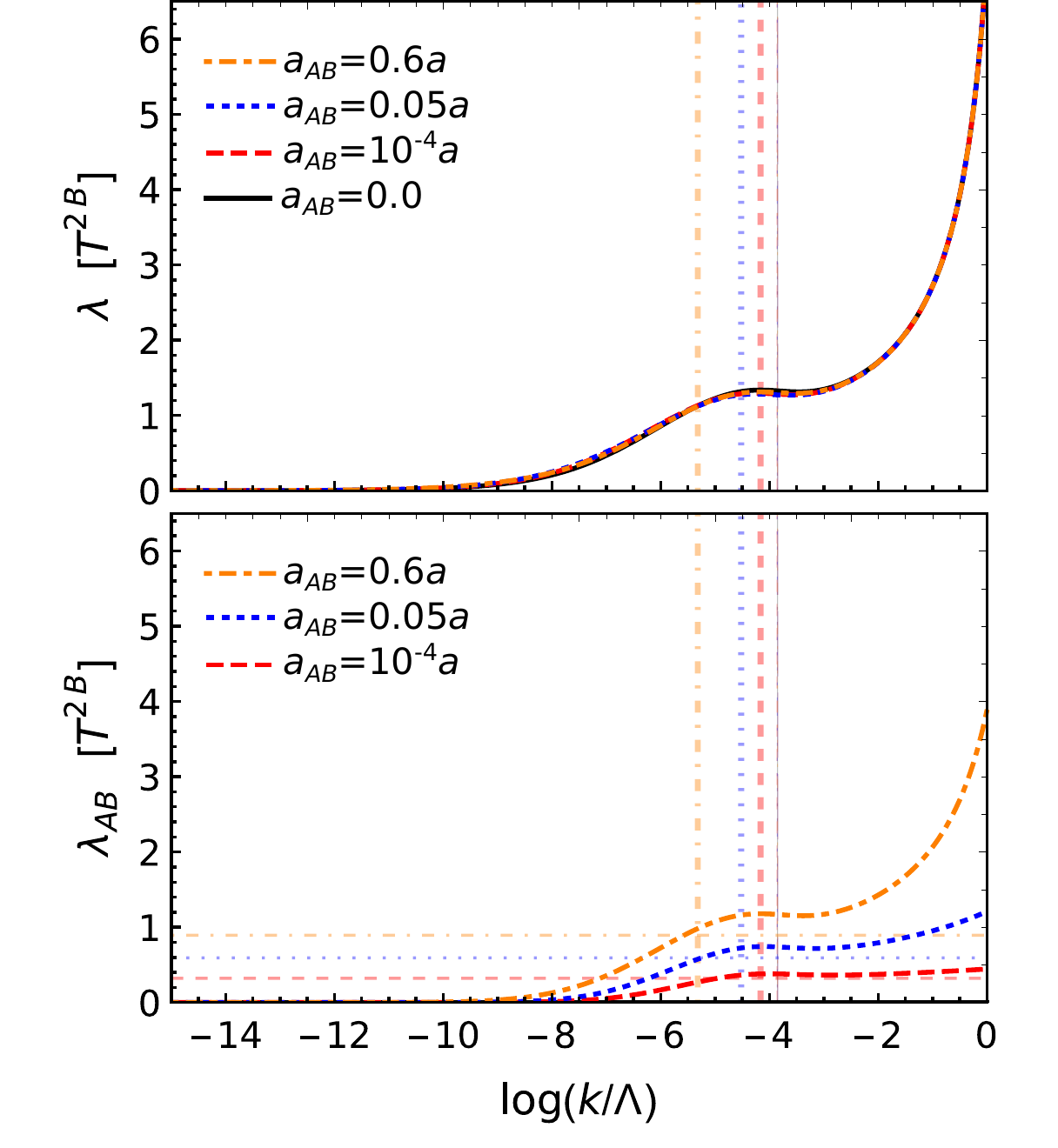}}
 \subfloat[Three dimensions]{\includegraphics[scale=0.65]{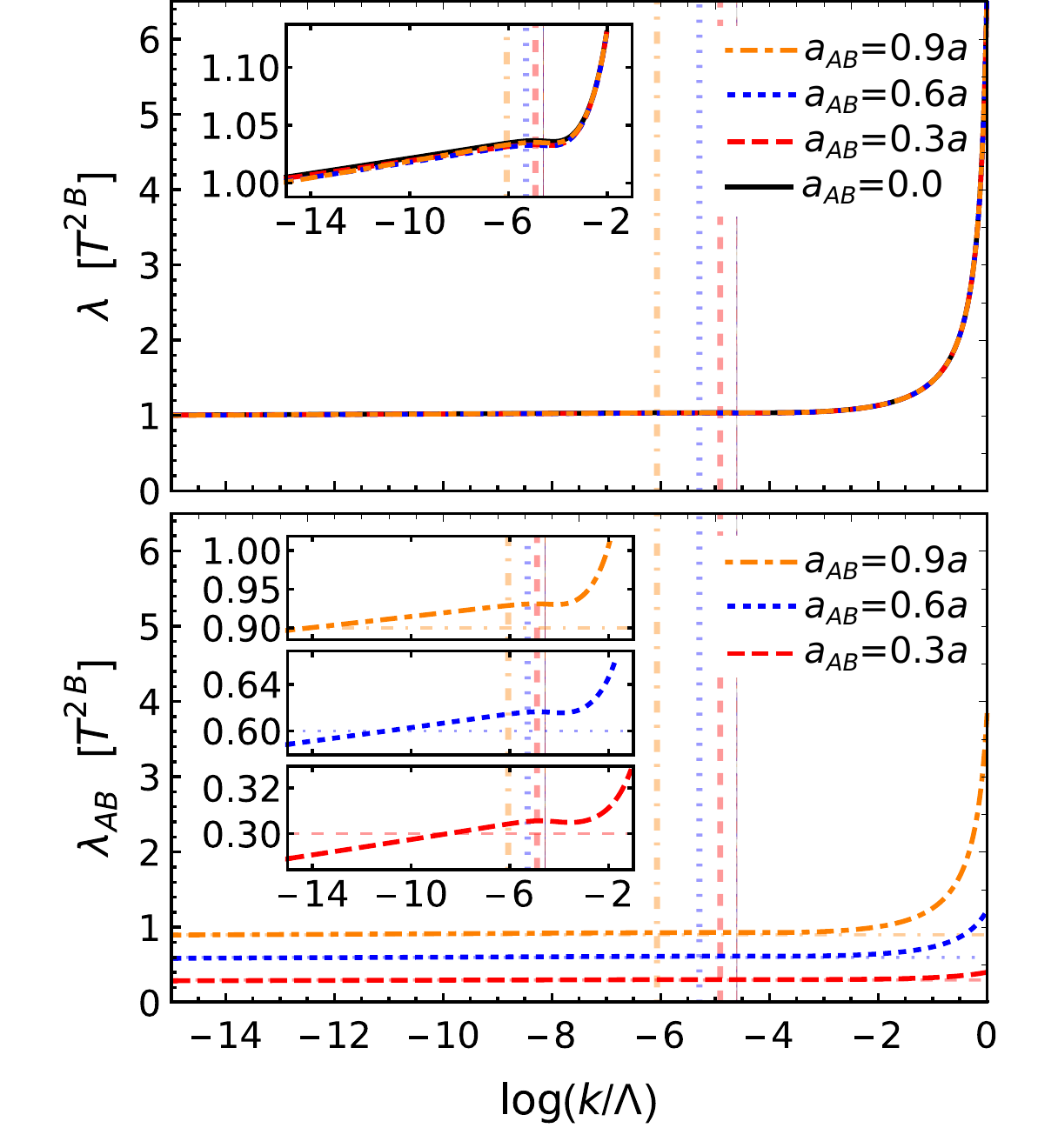}}
\caption{Flows of $\lambda$ and $\lambda_{AB}$ as a function of $k$  in two
and three dimensions for $m a^2 \mu=10^{-4}$. Both couplings are rescaled in terms of
the intra-species $T$-matrices defined in Eq.~(\ref{sec:FRG;sub:flow_eqs;eq:T2B}).
Thin and thick vertical lines denote $k_{h,+}$ and $k_{h,-}$, respectively. (Bottom) Solid horizontal lines correspond to the ratios
$T^\text{2B}/T^\text{2B}_{AB}$.}
\label{app:Flows;sub:Flows;fig:lambdas}
\end{figure*}

Fig.~\ref{app:Flows;sub:Flows;fig:lambdas} shows the flows of the interaction couplings
$\lambda$ and $\lambda_{AB}$. The intra-species coupling $\lambda$ is insensitive
to the inter-species interaction, with indistinguishable flows for different
values of $a_{AB}$. In contrast, $\lambda_{AB}$ depends strongly on $a_{AB}$, as expected.
As discussed in the main text, both $\lambda$ and $\lambda_{AB}$ vanish for $k\to 0$, 
with $\lambda$ vanishing as $k$ in two dimensions and logarithmically in three dimensions.
Consistent with our analysis for the ratio $\lambda_{AB}/\lambda$,
$\lambda_{AB}$ vanishes as $k^2$ in two dimensions and logarithmically in
three dimensions.

\begin{figure*}[t!]
\centering
 \subfloat[Two dimensions]{\includegraphics[scale=0.65]{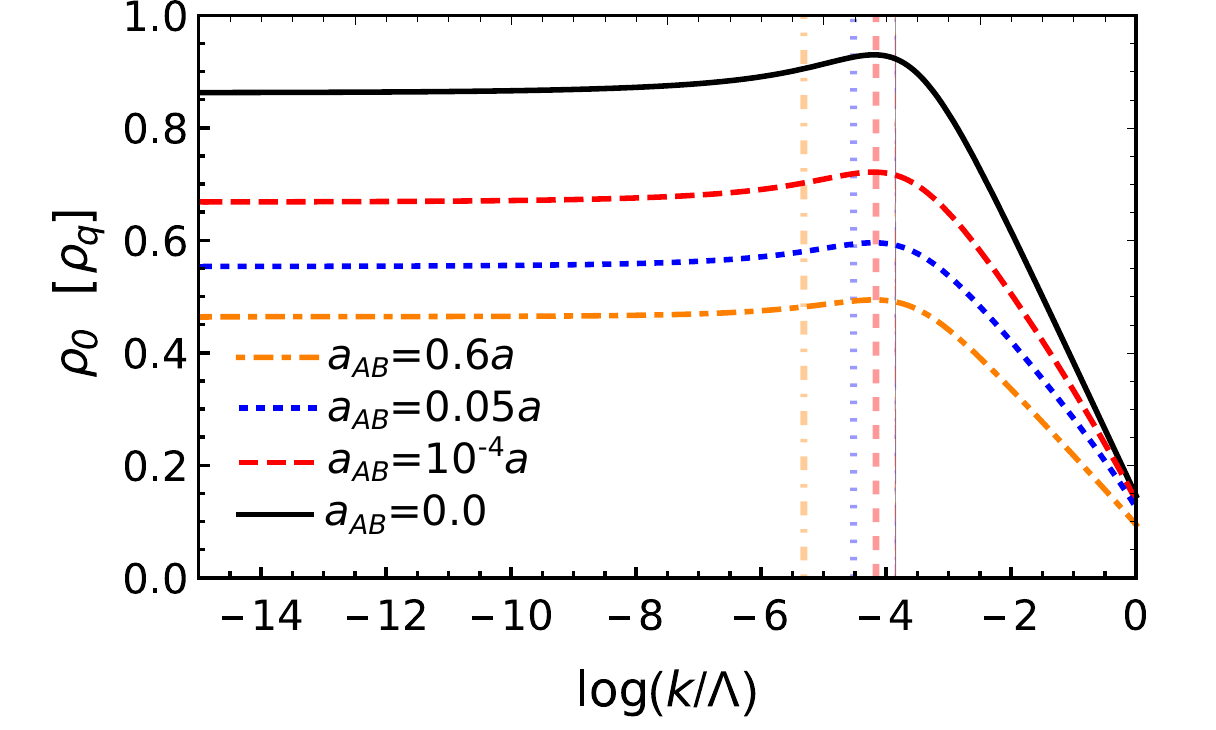}}
 \subfloat[Three dimensions]{\includegraphics[scale=0.65]{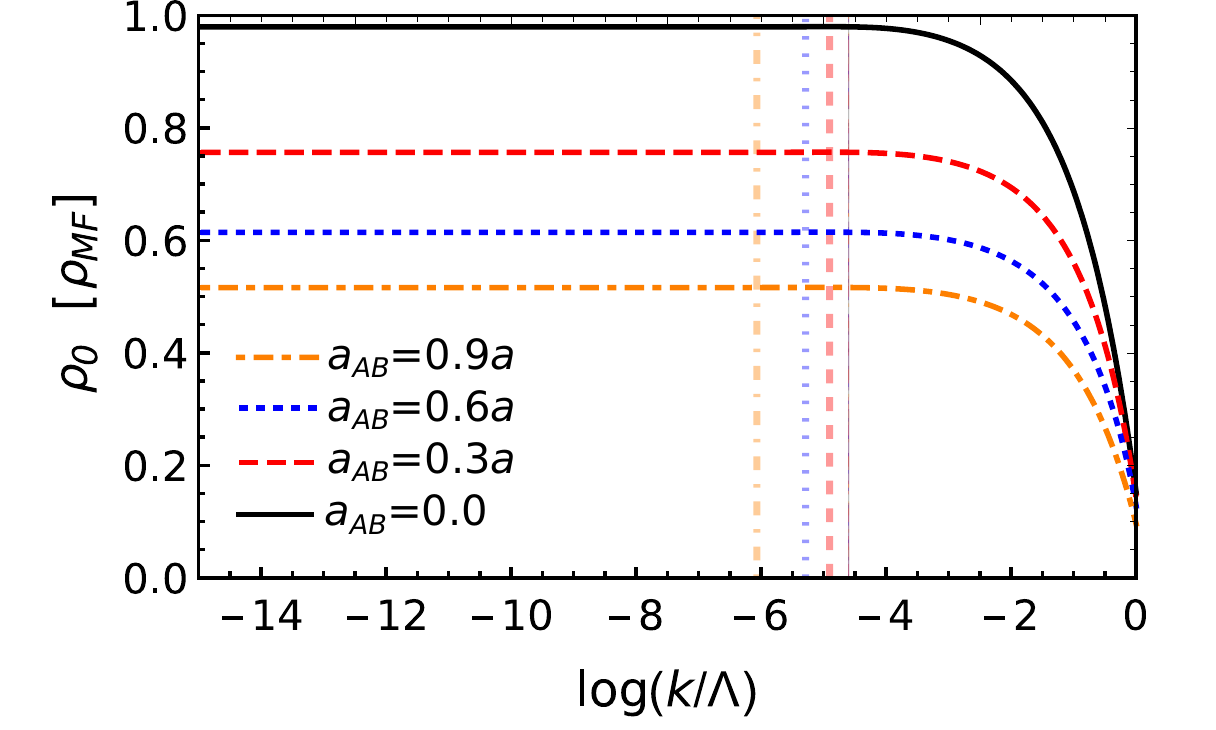}}
\caption{Flows of $\rho_0$ as a function of $k$  in two
and three dimensions for $m a^2 \mu=10^{-4}$.
In three dimensions $\rho_0$ is rescaled in terms of its MF value for
a one-component gas $\rho_\text{MF}=\mu m/4\pi a$. Similarly, in two dimensions $\rho_0$ is rescaled in terms of $\rho_q=\mu m \log(4/\mu m a^2e^{2\gamma_E+1})/4\pi$~\cite{mora_extension_2003}.
Thin and thick vertical lines denote $k_{h,+}$ and $k_{h,-}$, respectively.}
\label{app:Flows;sub:Flows;fig:rho0}
\end{figure*}

Finally, Fig.~\ref{app:Flows;sub:Flows;fig:rho0} shows flows of the condensate density $\rho_0$.
These are rescaled in terms of mean-field values for the one-component gas.
In all cases, $\rho_0$ converges to finite values for $k\to 0$, consistent with a finite condensate density.

\bibliography{biblio}

\end{document}